\documentclass[12pt,preprint]{aastex}

\slugcomment{Accepted for publication in the Astrophysical Journal Supplement Series}
\shorttitle{Analysis of Multicolor Photometry for sdB Stars} 
\shortauthors{P.-E. Tremblay, G. Fontaine, P. Brassard, P. Bergeron,
S.K. Randall} 

\newcommand{\gta}{\lower 0.5ex\hbox{$ \buildrel>\over\sim\ $}}
\newcommand{\lta}{\lower 0.5ex\hbox{$ \buildrel<\over\sim\ $}}
\newcommand{\Teff}{T_{\rm eff}}

\begin{document}

\title{A Quantitative Analysis of the Available Multicolor Photometry
  for Rapidly Pulsating Hot B Subdwarfs}

\author{P.-E. Tremblay, G. Fontaine, P. Brassard, P. Bergeron, and
  S.K. Randall} 

\affil{D\'epartement de Physique, Universit\'e de Montr\'eal, C.P.
                 6128, Succ. Centre-Ville, Montr\'eal, Qu\'ebec, Canada
                 H3C 3J7}

\email{tremblay@astro.umontreal.ca, fontaine@astro.umontreal.ca,
  brassard@astro.umontreal.ca, bergeron@astro.umontreal.ca,
  randall@astro.umontreal.ca}  

\begin{abstract}

We present a quantitative and homogeneous analysis of the broadband
multicolor photometric data sets gathered so far on rapidly pulsating
hot B subdwarf stars. This concerns seven distinct data sets related to
six different stars. Our analysis is carried out within the theoretical
framework developed by Randall et al., which includes full 
nonadiabatic effects. The goal of this analysis is partial mode
identification, i.e., the determination of the degree index $l$ of each
of the observed pulsation modes. We assume possible values of $l$ from 0
to 5 in our calculations. For each target star, we compute a specific 
model atmosphere and a specific pulsation model using estimates of the
atmospheric parameters coming from time-averaged optical spectroscopy. 
For every assumed value of $l$, we use a formal $\chi^2$ approach to
model the observed amplitude-wavelength distribution of each mode, and
we compute a quality-of-fit $Q$ probability to quantify the derived fit and to 
discriminate objectively between the various solutions. We find that no
completely convincing and unambiguous $l$ identification is possible on
the basis of the available data, although partial mode discrimination 
has been reached for 25 out of the 41 modes studied. A brief statistical 
study of these results suggests that a majority of the modes must have $l$
values of 0, 1, and 2, but also that modes with $l =4$ could very well
be present while modes with $l =3$ appear to be rarer. This is in line
with recent results showing that $l = 4$ modes in rapidly pulsating B
subdwarfs have a higher visibility in the optical domain than modes with
$l = 3$. Although somewhat disappointing in terms of mode
discrimination, our results still suggest that the full potential of 
multicolor photometry for $l$ identification in pulsating subdwarfs is
within reach. It will be a matter of gathering higher S/N ratio
observations than has been done up to now.

\end{abstract}


\section{INTRODUCTION}

It is well established that an analysis of multicolor photometric data,
whereby one compares oscillation amplitudes and/or phase differences in
two or more wavebands, can lead to partial mode identification in
pulsating stars (see, e.g., Heynderickx, Waelkens, \& Smeyers 1994). The
technique has been used successfully in the past to infer the values of
the degree index $l$ of pulsation modes for many types of oscillating
stars. To name just a few, let us mention that it has been applied to
$\delta$ Scuti stars (Garrido, Garc\'\i a-Lobo, \& Rodriguez 1990),
$\beta$ Cepheids (Cugier, Dziembowski, \& Pamyatnykh 1994), ZZ Ceti
white dwarfs (Fontaine et al. 1996), $\gamma$ Doradus stars (Breger et
al. 1997), and Slowly Pulsating main sequence B stars (Dupret et
al. 2003). Attempts to understand the observed period spectrum on the
basis of multicolor photometry have also been made for EC 14026 stars,
first by Koen (1998) and more recently by Jeffery et al. (2004;
2005). The former study gives a qualitative interpretation of 
the periods observed for KPD 2109+4401 based on the theory of Watson
(1988), while the latter assert to have provisionally identified or
constrained the $l$ values for some of the modes detected for the fast
oscillators KPD 2109+4401, HS 0039+4302, and PG 0014+067. In these
cases, the $l$ identification is based on a qualitative comparison with
amplitude ratios computed by Ramachandran, Jeffery, \& Townsend (2004)
in the adiabatic approximation for representative models.

A theoretical framework for the quantitative exploitation of multicolor
photometry for pulsating sdB stars has been put forward recently by some of
us (Randall et al. 2005). In that paper, the potential of the technique
as applied to both types of pulsating sdB stars (the short-period
$p$-mode pulsators of the EC 14026 type and the slowly oscillating
$g$-mode variables of the PG 1716 type) has been explored in
detail. The method of Randall et al. (2005) features a full nonadiabatic
treatment of the atmospheric layers and uses a designated model
atmosphere code which automatically incorporates the wavelength
dependence of the limb darkening --- thus avoiding the need for
approximate parameterized limb darkening coefficients as used in most
other multicolor photometric studies with the notable exception of
Ramachandran et al. (2004). Of central importance, the method
of Randall et al. (2005) has led to the first (and, so far, only)
unambiguous determination of the $l$ index of a pulsation mode in a
oscillating sdB star using multicolor photometry, thus demonstrating the
feasibility of the approach.  

Indeed, at the time of the writing of the Randall et al. paper, the
best available amplitude estimates of oscillation modes in a pulsating
EC 14026 star were those based on the superb $u'g'r'$ data set gathered
by Jeffery et al. (2004) on KPD 2109+4401 ($V$ = 13.38) with {\sl
  ULTRACAM} mounted on the 4.2-m William Herschel Telescope
(WHT). Considering the largest amplitude mode reported with a period of
182.42 s, and adopting at face value the three amplitudes and their
quoted uncertainties ($u' = 8.87 \pm 0.04$ mmag, $g' = 6.58 \pm 0.04$
mmag, $r' = 6.15 \pm 0.04$ mmag), Randall et al. (2005) were able to
demonstrate that this mode must be a radial mode ($l$ = 0) as first
indicated by Jeffery et al. (2004).

The quality of these unique observations has remained unsurpassed so far,  
but there are nevertheless other available multicolor data sets that
certainly deserve to be analyzed in the same fashion now that there is a
proven theoretical framework to do that. While it is known that it
is generally difficult to discriminate between modes with $l$ = 0, 1, or
2 on the basis of multicolor photometry for EC 14026 stars (Ramachandran
et al. 2004; Randall et al. 2005), one may hope to distinguish between
modes of that group from modes with $l$ = 3 or $l$ = 4 or possibly $l$ =
5. Our survey of the published work in the field has revealed the
following available data sets 1) the $UBVR$ photometry of Koen (1998;
herafter K98) on KPD 2109+4401, 2) the $UBV$ photometry of Silvotti et
al. (2000; herafter S00) on PG 1628+563B, 3) the {\sl BUSCA} photometry
of Falter et al. (2003; hereafter F03) on PG 1605+072, 4) the
$u'g'r'$/{\sl ULTRACAM} photometry of Jeffery et al. (2004; hereafter
J04) on HS 0039+4309, 5) the $u'g'r'$/{\sl ULTRACAM} photometry of
Jeffery et al. (2004) on KPD 2109+4401, 6) the Str\"omgren photometry of
Oreiro et al. (2005; hereafter O05) on BAL 090100001, 7) the $UBVR$
photometry of Baran et al. (2005; hereafter B05) on BAL 090100001, 8) the
$u'g'r'$/{\sl ULTRACAM} photometry of Jeffery et al. (2005; hereafter
J05) on PG 0014+067, and 9) the $u'g'r'$/{\sl ULTRACAM} photometry of
Aerts et al. (2006) on SDSS J171722.08+58055.8. Except for cases 5) and
9) (see below), we analyze all of the other data sets in the present
paper.

Partial mode identification (the identification of the degree index $l$)
through multicolor photometry is a worthwhile venture in itself, but it
has become particularly important for EC 14026 pulsators now that
specific asteroseismic models have been proposed for a few of them. 
Indeed, by combining the forward method with high S/N ratio spectroscopy, 
it has been possible to carry out complete asteroseismological analyses
for four EC 14026 pulsators so far: PG 0014+067 (Brassard et al. 2001),
PG 1047+003 (Charpinet et al. 2003), PG 1219+534 (Charpinet et
al. 2005a), and Feige 48 (Charpinet et al. 2005b). The ultimate product
of these analyses is the determination of the global structural
properties of the stars, but complete mode identification (i.e., a
determination of the radial order $k$ and the degree index $l$ of each
pulsation mode) is a byproduct of the method. Multicolor photometry can
thus be used as testing grounds for the proposed seismic models by
checking if the $l$ assignments are correct. Of the four EC 14026
pulsators with a proposed specific seismic model, only PG 0014+067 has
been observed so far in several filters simultaneoulsy (J05). We will
discuss this interesting case below as part of our analysis of that data
set.

The main purpose of the paper is to present a quantitative and homogeneous
analysis based on the method of Randall et al. (2005) for each of the
available multicolor data sets that we found in the published literature
on pulsating EC 14026 stars. Considerable observational efforts went
into the gathering of these data sets and we feel that it is most
worthwhile to attempt extracting the maximum information from them. 
Partial mode identification is the goal. We note that the exact same
approach has been used recently by Aerts et al. (2006) for analyzing their 
{\sl ULTRACAM}/WHT data of the very faint EC 14026 pulsator
SDSS J171722.08+58055.8 ($B$ $\simeq$ 16.7), so it is not necessary to
reconsider their data here. Unfortunately, in that case, the
identification of $l$ was not possible for the two pulsation modes 
uncovered because the uncertainties on the observed amplitudes turned
out to be too large due to insufficient S/N ratio. In contrast, some
interesting mode identifications appear possible for the Jeffery et
al. (2004) data set on KPD 2109+4401 (beyond that for the 182.42 s mode
discussed in Randall et al. 2005), but we defer our analysis to a
separate publication that will include unpublished additional
observations that we gathered at the Canada-France-Hawaii 3.6-m
Telescope on that star. We note also that two-color ($U$ and $R$)
photometry has been gathered and analyzed for two PG 1716 sdB pulsators
in order to constrain the $l$ index (see the paper of Randall et
al. 2006a on PG 1627+017 and that of Randall et al. 2006b on PG
1338+481).

\section{METHOD}

\subsection{The computations of theoretical amplitudes}

To analyze multicolor data for pulsating sdB stars, we follow the
modeling method developed by Randall et al. (2005). While we refer the
reader to that paper for more information, we include here a brief 
recap of the most relevant aspects for convenience. The
technique employed is based on the well-known fact that the observed 
wavelength dependence of an oscillation's amplitude and phase bears the
signature of the mode's degree index $l$ and its period, as well as the star's
atmospheric parameters, the intrinsic amplitude of the periodicity, and
its viewing aspect. To first order, the influence of the latter two
(unknown) parameters can be eliminated by computing the ratio of pulsational
amplitudes (and the difference between phases) as measured in two or
more bandpasses. Given a target's effective temperature and surface
gravity, the observed mode's degree index can thus, in principle, be
inferred from multicolor photometry, leading to partial mode
identification.
 
The method proposed by Randall et al. (2005) incorporates a full
nonadiabatic description of the atmospheric layers in the computations
of theoretical pulsation observables. As in main sequence stars (e.g.,
Dupret et al. 2003), such description is found to be quite important 
in sdB stars as well. In particular, while the predicted phase shifts
between various bandpasses generally remain quite small --- they 
would all be identical to zero in the adiabatic approximation --- Randall et
al. (2005) found that amplitude ratios cannot, in that approximation,
be computed with enough accuracy for quantitative studies. Furthermore,
they found that the amplitude ratios do depend sensitively on the
atmospheric parameters of the target, so for the purposes of our present
analysis, we need to compute a specific detailed model atmosphere as
well as a full pulsation model for each one of our targets. 

For each value of $l$, the brightness variation expected across the
visible disk during a pulsation cycle can be expressed in terms of 
temperature, radius, and surface gravity perturbations to the emergent
flux. These in turn are dependent on quantities obtainable from model
atmospheres and nonadiabatic pulsation theory, as well as on the period
of the mode in question (see, e.g., equations 34 $-$ 38 of Randall et
al. 2005). The model atmosphere parameters are made up of the
logarithmic derivative of the emergent flux with respect to the
effective temperature / surface gravity and the weighted limb darkening
integral together with its derivatives. Initially computed as
monochromatic quantities ($\alpha_{T\nu}$, $\alpha_{g\nu}$, $b_{l\nu}$,
$b_{l\nu ,T}$, and $b_{l\nu,g}$ in the notation of Randall et al. 2005) 
from a specially modified sdB atmosphere code, they are subsequently
integrated over the bandpasses of interest (the quantities
$\alpha_{Tx}$, $\alpha_{gx}$, $b_{lx}$, $b_{lx,T}$, and $b_{lx,g}$,
where $x$ symbolizes the appropriate bandpass, again in the notation of
Randall et al. 2005) to allow for comparison with observations. 

The second set of parameters includes the nonadiabatic quantities $R$
and $\Psi_T$, which are related to an oscillation's departure from
adiabacity in amplitude and phase respectively. Together with the
adiabatic gradient, $\nabla_{ad}$, they are computed on the basis of the
so-called ``second generation'' static stellar models described in
Charpinet et al. (2001). These full models are submitted to adiabatic and
nonadiabatic pulsation calculations that yield, among other things, the
relative behaviour of the radius (gravity) and temperature perturbations
in the stellar atmosphere for each mode considered. Integrated over the
atmospheric layers contributing most to the emergent flux (taken to lie
at optical depths $\tau$ = 0.1$-$10), the ratio of the two
perturbations' moduli and their phase lag yield the atmosphere-averaged
values of $R$ and $\Psi_T$ describing the departure from adiabacity of
the observed brightness variations for each mode. The atmospheric value of
$\nabla_{ad}$ is obtained in a similar way.

For each reported mode in each observed target, we calculated the
predicted amplitude ratios and phase differences from the available
bandpasses for values of the degree index from $l$ = 0 to $l$ =
5. Since, as it turns out, there is very little information content in
the phase shifts (the predicted differences are indeed too small in all
cases of interest to be detectable at the accuracy achieved in the
available data sets), we concentrate solely in what follows on comparing
predicted and observed amplitudes. To provide a quantitative framework
for such a comparison, we contrast for each mode the predicted
multicolor amplitudes with those observed using a $\chi^2$ minimization
procedure following Fontaine et al. (1996).  For every degree
index $l$, the theoretical amplitudes $a_{theo}^i$ in each of the $N$
available bandpasses $i$ are multiplied by a scale factor $f_l$ chosen
in such a way as to minimize 
\begin{equation}
\chi^2(l)=\sum_{i=1}^N\displaystyle\bigg(\frac{f_l
  a_{theo}^i-a_{obs}^i}{\sigma^{i\
}}\displaystyle\bigg)^2,
\end{equation}
where $a_{obs}^i$ is the amplitude observed in a given waveband and
$\sigma^i$ is the error on the measurement. Compared to the standard
normalization of all amplitudes to one particular waveband, this is a
more objective way of determining the quality of a match, since the data
from all bandpasses are treated on the same footing.

Finally, since our procedure is a standard $\chi^2$ approach, it is
possible, following Fontaine et al. (1996), to quantify the value of a
given $\chi^2$ solution by computing explicitly the quality-of-fit $Q$
as described in Press et al. (1986). The quantity $Q$ depends on the
value of $\chi^2$ for each solution and the number of degrees of freedom
($N$ fitted points minus the free parameter $f_l$ gives $N-1$ degrees of
freedom in the present case). We adopt the canonical notion suggested by
Press et al. (1986) that a fit is acceptable if its quality-of-fit
$Q>0.001$. For a system with a single degree of freedom, this value of
$Q$ corresponds to the case where the predicted amplitudes fall
simultaneoulsy in all bandpasses of interest within 3$\sigma^i$. Hence,
we shall use the criterion $Q>0.001$ to determine if a given $\chi^2$
fit is acceptable and to discriminate quantitatively betweeen possible
$l$ solutions.

\subsection{The available data sets}

The vital characteristics of each of the seven data sets used in this
study are summarized in Table 1. The first column gives the name of the
EC 14026 pulsator, the second one gives its $V$ magnitude ($B$ if $V$ is
unavailable), the third one lists the total length of the data set, the
fourth column refers to the bandpasses of interest, the fifth one
indicates the instrument used, and this is followed by the
identification of the telescope used and its location, and finally a reference
to the original paper is given.

In the calculations of theoretical amplitudes for each target star, it
is necessary to specify the atmospheric parameters of the star, log $g$
and $T_{\rm eff}$. To do this, we rely on the recent and ongoing
spectroscopic efforts of Green, Fontaine, \& Chayer (2006, in
preparation) to derive reliable and homogeneous atmospheric parameters
for a large sample of sdB stars. For each of our target stars, we used
estimates for the atmospheric parameters coming from that study. These
are given in Table 2.

In addition, we indicate in the same table the three ingredients that we
used in the convolution process transforming monochromatic quantities
into waveband-integrated quantities suitable for comparison with the
observational data. These include the transmission curve of each filter,
the response curve of the detector, and the extinction curve of the
site. Tests indicate that, by far, the transmission curves of the filters
are the most important component, and that, at the other extreme, the
choice of the extinction curve is a sophistication that has little
impact for sites located at roughly the same altitude. In this
connection, we picked the atmospheric transparency curve of a
representative medium altitude site (in this case, Kitt Peak National
Observatory), a choice that should be appropriate for all but perhaps
Mt. Suhora Observatory in Poland located at an altitude of about 1000 m.
Further tests indicate that the largest sources of uncertainty for the
computed amplitudes are associated with the uncertainties on $\Teff$ and
log $g$, assuming that the filters used by the observers have indeed the
claimed transmission properties. For typical uncertainties on the
atmospheric parameters of $\Delta T_{\rm eff}$ = $\pm$ 400 K and $\Delta
\log g$ = $\pm$ 0.05 (this excludes possible systematic effects
associated with the model atmospheres used by Green et al. 2006), the
real effects on the theoretical amplitudes remain quite small and
generally not significant enough to affect the implications for mode
identification (given the observational uncertainties on the observed
amplitudes). 

For each data set, we generally accepted the data, i.e., the observed 
multicolor amplitudes and their associated uncertainties at face value. 
This implicitly implies that each reported mode was fully resolved and
that its amplitudes were free of spectral contamination from neighborhing
peaks in the Fourier domain. This is an important issue: if a mode is
not sufficiently ``monochromatic'' in frequency, attempts to model its
$l$ signature as a function of wavelength are usually frustrated. One
exception to this is the case of an unresolved frequency multiplet due
to rotational splitting. In that case, since the amplitudes do not
depend on the azimuthal order $m$, the modeling remains valid for such 
an unresolved multiplet in a slowly rotating star.

An equally important concern is the question of the uncertainties on the 
observed amplitudes. Since our approach is based on a $\chi^2$
statistics, it is absolutely essential to have realistic and accurate
estimates of these uncertainties. The quality of the fit, as measured by
the quantity $Q$, and which gives the discriminatory power of the
method, depends directly on the values of these uncertainties. We found,
in this connection, that our failure sometimes to model in an
acceptable way a given pulsation mode could be traced back to the fact
that the errors on the amplitudes were probably underestimated. 

\section{RESULTS}

\subsection{KPD 2109+4401 (K98)}

The paper of K98 is the first one reporting on multicolor photometry for
a rapidly pulsating sdB star. It was a key source of motivation and
inspiration for the theoretical work that we presented in Randall et
al. (2005). In K98, the author reported, among other things, on some 27
h of $UBVR$ photometry on KPD 2109+4401 ($V$ = 13.38) gathered with a
4-channel photometer mounted on the McDonald Observatory's 2.1-m Struve
Telescope. Seven distinct pulsation modes were uncovered, and $UBVR$
amplitudes and phases were provided for each one of them. The author,
however, was clearly reluctant to provide estimates of the uncertainties
on the amplitudes and phases, being quite aware, as he argued, that
formal errors coming from least-squares fits of light curves tend to be
unreliable. Reading between the lines, and using the information on the
relative efficiency of each bandpass of the Steining photometer that we
had from previous work (Fontaine et al. 1996), we established that
reasonable estimates of the observational uncertainties on the
amplitudes would be $\sigma(U)$ = 0.105 mmag, $\sigma(B)$ = 0.105 mmag,
$\sigma(V)$ = 0.085 mmag, and $\sigma(R)$ = 0.155 mmag, irrespective of
the amplitude of a given mode. 

Table 3 summarizes the results of our modeling effort for the K98 data
set.\footnote{Note that we explicitly used the transmission curves of
the $UBVR$ filters of the Steining photometer as published in Robinson
et al. (1995) for this analysis. These are similar but not exactly the
same as the standard Johnson-Cousins filters.} For each observed mode,
the table gives a block of data. The first line in each block gives the
period of the mode of interest, and this is followed by the observed
amplitude and its uncertainty in each of the available bandpass. The
next line gives the best-fit theoretical amplitude in each of the
bandpass for an assumed degree index $l$ = 0, and this is followed by
the value of $\chi^2$ obtained in the minimization procedure (eq. 1),
and the value of the quality-of-fit $Q$. The next five lines show
similar data, but for assumed values of $l$ from 1 to 5, respectively.
  
Figure 1 is a graphical representation of our results for the 182.42 s
mode (the first one listed in Table 3 and the second largest amplitude
one in the K98 data). The observed behavior of that mode is well
modeled for values of $l = 0$, $l = 1$, or $l = 2$ as can be seen in
the figure. The plot is consistent, of course, with the values of $Q$
all larger than 0.001 for those three possibilities, while it also
clearly shows the poor fits ($Q \ll 0.001$; see Table 3) obtained for
models assuming $l = 3$, $l = 4$, or $l = 5$. We thus find that, while
the uncertainties on the observed amplitudes are too large for us to
discriminate between the values of $l$ from 0 to 2, we can safely
discard the possibilities that the 182.42 s mode in KPD 2109+4401 has a
degree index of 3, 4, or 5. We note that this result is consistent with
the determination of $l = 0$ proposed by Randall et al. (2005) for that
mode on the basis of the the much higher S/N ratio data of Jeffery et
al. (2004).

The highest amplitude mode in the K98 data is the 196.31 s pulsation
which, interestingly enough but not uncommonly in sdB pulsators, showed
smaller amplitudes than the 182.42 s mode when KPD 2109+4401 was observed by
Jeffery et al. (2004). Our modeling of the K98 data shows a very similar
situation to the previous mode as can be seen in Figure 2. Again, we
cannot formally discriminate between $l = 0$, $l = 1$, or $l = 2$, but
we can safely exclude the possibilities that the 196.31 s mode in KPD
2109+4401 has a value of $l = 3$, $l = 4$, or $l = 5$. The very small
values of $Q$ associated with these solutions (Table 3) certainly confirm
this conclusion.

The most interesting case in the K98 data is that of the 198.19 s mode
(the third largest amplitude mode), which shows an amplitude-wavelength
behavior characteristic of a $l = 4$ mode as can be appreciated in
Figure 3. In fact, formally speaking, the $l = 4$ solution is the only
one acceptable, with a value of $Q > 0.001$ in Table 3. While this
remains highly suggestive, we caution that one should not jump too
hastily to the conclusion that mode discrimination has been achieved
beyond any doubt for that mode. The reason for our cautionary remark is
that the next best fit, the one corresponding to $l = 2$ and $Q= 2.77
\times 10^{-4}$ in Table 3, does not show a quality-of-fit value that is
much smaller than the passage criterion of $Q = 0.001$. For instance, if
we were to increase the uncertainties on the observed amplitudes by some
10\%, the solution with $l = 2$ would become formally acceptable. So
here is a case where realistic and accurate estimates of the amplitude
uncertainties become critical.

For the other modes uncovered in K98, we formally constrain the value of
$l$ to 0, 1, 2, or 4 for the 184.72 s pulsation, to 2 or 4 for the
184.75 s pulsation, and to 1, 2, or 4 for the 191.85 s pulsation. No
mode discrimination was possible for the lowest amplitude mode at 196.69
s, and all values of $l$ from 0 to 5 provide acceptable fits. We note
that the provisionary $l$ assignments made by Jeffery et al. (2004) on
the basis of their {\sl ULTRACAM}/WHT observations are consistent with
our more quantitative results here for all modes, except perhaps for the
198.19 s pulsation for which we may have isolated a unique value at $l =
4$, although, as discussed just above, the $l = 2$ solution (which is the
suggestion of Jeffery et al. 2004 for that mode) should not be dismissed
too hastily. 

\subsection{PG 1618+563B (S00)}

Amplitude and phase data were reported for two modes (139.3 s and 143.9
s) observed in PG 1618+563B ($V$ = 13.52) by S00. Those were derived
from $UBV$ observations using the 3-channel Tromso photometer attached
to the 2.5-m NOT Telescope. Unfortunately, only 1.5 h of data were
gathered and, consequently, the S/N ratio was not large enough to allow
$l$ index discrimination. Our results are summarized in Table 4 (the
format is similar to that of Table 3) and clearly show that all possible
$l$ values from 0 to 5 provide acceptable fits for both modes. Figure 4
gives a graphical representation of our results for the 139.3 s mode.

\subsection{PG 1605+072 (F03)}

The relatively bright ($V$ = 12.92), large-amplitude sdB pulsator PG
1605+072 has been observed by F03 who provided 12.3 h of {\sl BUSCA}
photometry, a 4-channel system allowing the simultaneous observations in
bandpasses centered on 3600 \AA ($UV$), 4800 \AA ($B$), 6300 \AA ($R$),
and 8000 \AA ($NIR$). These data were gathered at the 2.2-m telescope at
Calar Alto. F03 isolated 11 modes in their light curves and provided
measurements of amplitudes and phases in their Tables 2 and 3. However,
only 5 of those modes have simultaneous measurements in all the 4 {\sl
BUSCA} bandpasses. We have restricted our analysis to these 5 modes.

In our modeling effort, we were careful to convolve the transmission
curves of the 4 filters with the response curve of the {\sl BUSCA}
instrument as provided to us by O. Cordes (2005, private
communication). We also point out that the estimates of the atmospheric
parameters that we used for PG 1605+072 (see Table 2) are remarkably
consistent with the independent values obtained by Heber, Reid, \&
Werner (1999), which are $T_{\rm eff}$ = 32,300 $\pm$ 300 K and log $g$ =
5.25 $\pm$ 0.05 dex.  

Despite careful modeling, we were unable to reproduce satisfactorily the
amplitude-wavelength behaviors of all 5 modes when using the amplitudes
and their uncertainties given in Tables 2 and 3 of F03. In all 5 cases,
and for all assumed values of $l$ from 0 to 5, we obtained unacceptable
quality-of-fit values of $Q \ll 0.001$. While we cannot completely rule
out the possibility that our modeling is inadequate for this particular
star, we believe instead that this problem is related to the fact that
the amplitude uncertainties quoted by F03 most likely underestimate the
true errors. In fact, the authors of F03 seem to be aware of that and
provide two other different estimates of the amplitude uncertainties.
For instance, they explicitly write that the quoted uncertainties on the
reported amplitudes in their Tables 2 and 3 are formal fit errors from a
least-squares sine fit procedure, while the (much larger) uncertainties
shown in their Figure 5 for 4 different modes have been estimated in
another way. Furthermore, F03 write in their subsection 3.2
that the 1$\sigma$ noise level in their photometry corresponds to 1.52
mmag in the $UV$ bandpass, 1.53 mmag in $B$, 1.12 mmag in $R$, and 1.37
mmag in $NIR$, which is clearly at odds with the much smaller formal
fit errors reported in their tables.

We used these last figures to derive more realistic estimates of the
amplitude uncertainties for the F03 data set. To this end, we adopted the
recipe of Montgomery \& O'Donoghue (1999), which suggests that the
amplitude uncertainty should be equal to about 0.8 times the 1$\sigma$
noise level, irrespective of the actual amplitude of a mode. This leads
to $\sigma(UV)$ = 1.22 mmag, $\sigma(B)$ = 1.22 mmag, $\sigma(R)$ =
0.90 mmag, and $\sigma(NIR)$ = 1.10 mmag. We note that these correspond
to values 10 to 25 times larger than the formal fit errors quoted by F03
in their Tables 2 and 3, a huge difference perhaps in line with the
reluctance shown by K98 to quote realistic errors for this type of
data. While the approach of Montgomery \& O'Donoghue (1999) is
conservative, we believe that it leads to more realistic estimates of
the amplitude uncertainties in this case, and we have consequently redone
our analysis of the F03 data set with these revised values.
   
Our results are summarized in Table 5. Partial mode discrimination is
possible for the three highest amplitude modes considered, and we show
the corresponding fits in Figures 5, 6, and 7. The best result is
obtained for the largest amplitude mode observed by F03 (481.75 s), for
which, according to our analysis, the $l$ value is equal to either 0, 1,
or 2. 

\subsection{HS 0039+4302 (J04)}

HS 0039+4302 has been observed by J04 using the 3-channel {\sl ULTRACAM}
camera attached to the 4.2-m William Herschel Telescope. Some 16.2 h of
$u'g'r'$ Sloan photometry has been acquired on this relatively faint EC
14026 pulsator at $V$ = 15.5. Six modes with $u'g'r'$ amplitudes have
been isolated by the authors. While they have quoted an $average$
amplitude uncertainty of 0.10 mmag for the 3 bandpasses of interest for
all the 6 modes in their Table 3, we can use the results of J05 and
Aerts et al. (2006) based on the same detector/telescope combination to
infer wavelength-dependent amplitude uncertainties of $\sigma(u')$ =
0.156 mmag, $\sigma(g')$ = 0.063 mmag, and $\sigma(r')$ = 0.081
mmag. These are the values that we used in our modeling of the 6
different modes. Note that during this modeling exercise, we convolved
the transmission curves of the Sloan filters $u'$, $g'$, and $r'$ with
the response curves of the CCD's used in {\sl ULTRACAM} kindly provided
to us by V. Dhillon (2005, private communication).

The results of our analysis for HS 0039+4302 are summarized in Table
6. They are interesting in that they show, despite the relative
faintness of the target, that partial mode identification is possible. Except
for the 134.44 s mode (the lowest amplitude one) for which mode
discrimination is not feasible, partial identification is indeed
possible. For instance, Figures 8, 9, and 10, corresponding to the
three largest amplitude modes, indicate that these pulsations have values of
the degree index $l$ of either 0, 1, or 2. We note that the preliminary
$l$ assignments made by J04 for the 6 modes they observed in HS
0039+4302 are consistent with our quantitative results.

\subsection{BAL 090100001 (O05)}

Oreiro et al. (2004) reported the discovery of short-period luminosity
variations in BAL 090100001, the brightest and largest amplitude EC 14026
pulsator so far discovered. From a strictly observational point of view,
these characteristics make it an ideal target to attempt multicolor
photometry. Hence, less than a year after the initial discovery of the
variability of BAL 090100001, O05 presented follow-up observations,
including 9.3 h of simultaneous Str\"omgren photometry on that
star. This data set was gathered with the dedicated Str\"omgren
photometer attached to the 0.9-m telescope at the Sierra Nevada
Observatory. 

Our attempts to model the amplitude-wavelength behaviors of the two
pulsation modes reported by O05 (see their Table 5) were frustrated and
no acceptable fits were found for any of the assumed $l$ values from 0
to 5. Table 7 summarizes our results and clearly indicates that our
quality-of-fit values $Q$ are completely unacceptable. Figure 11
illustrates graphically the situation for the largest amplitude mode
(356.3 s). Again, as in the case of PG 1605+072 discussed above, we
cannot exclude completely the possibility that our modeling is
inadequate for BAL 090100001, but we rather strongly suspect that the 
quoted amplitude uncertainties in Table 5 of O05 (most likely formal fit
errors) underestimate largely the true errors. Unfortunately, unlike the
case of PG 1605+072 in F03, not enough information is given in O05 that
would have allowed us to obtain perhaps more realistic estimates of the
amplitude uncertainties. Thus, we must conclude that we cannot model
adequately the data of O05 as presented by them. 

\subsection{BAL 090100001 (B05)}

BAL 090100001 was also observed by B05 who, in a major effort, gathered
more than 126 h of $UBVR$ photometry on the 0.6-m telescope of the
Mt.~Suhora Observatory in Poland. The authors uncovered many
periodicities, including not only short-period pulsations characteristic
of EC 14026 stars, but also long-period oscillations most likely due to
$g$-mode pulsations as found in the PG 1716 stars (and see also O05). We
have concentrated our modeling effort on the 9 short-period modes
(labelled $f_1$ through $f_9$ in B05) with reliable amplitude
determinations in all four bandpasses. We note that the authors 
provided a very careful frequency analysis and that they adopted the
approach of Montgomery \& O'Donoghue (1999) for estimating the
uncertainties on the reported amplitudes and phases. For the amplitudes,
this leads to $\sigma(U)$ = 0.25 mmag, $\sigma(B)$ = 0.18 mmag,
$\sigma(V)$ = 0.16 mmag, and $\sigma(R)$ = 0.17 mmag, irrespective of
the actual amplitudes of a given mode.   

The light curve of BAL 090100001, as observed by B05, is dominated by a
very large amplitude mode (the largest ever observed so far in a sdB
pulsator) with a period of 356.19 s. Its color amplitudes are $U$ =
75.23 mmag, $B$ = 57.71 mmag, $V$ = 53.34 mmag, and $R$ = 50.26 mmag
(see Table 1 of B05). Not surprisingly, B05 also report the detection of
the first and second harmonic of that mode, the first with an amplitude
larger than 6 of the 9 modes we retained for analysis. The dominant mode
appears to be isolated, whereas the three next largest ones (354.20 s,
354.01 s, and 353.81 s) appear to form an almost perfectly symmetric
triplet in frequency space, leading to the suggestion that this triplet
could be due to rotational splitting. Considering that the main mode
does not show an equivalent multiplet structure, this led B05 to suggest
that the 356.19 s mode could be a $l = 0$ mode, while the triplet would
correspond to a rotationally-split $l = 1$ pulsation. It is obviously of
high interest to verify if these sensible suggestions could be proven true.

To model the data of B05, we adopted the atmospheric parameters given in
Table 2 ($T_{\rm eff}$ = 29,810 K and log $g$ = 5.58), and we convolved
our basic monochromatic quantities with the standard transmission curves
of the Johnson/Cousins filters, the KPNO extinction curve, and a gray
response for the CCD detector since, in the latter case, we had no
information as to the exact response curve of the detector. The results
of our effort for the 9 modes retained are summarized in Table 8. In
addition, we include here Figures 12, 13, 14, and 15 that refer,
respectively, to the main mode and the triplet mentioned above. The
other modes have relatively low amplitudes and it becomes increasingly
difficult to discrimate between the possible values of $l$ for them. 

Figure 12 illustrates our results for the main mode. Formally speaking,
all model fits shown in the figure for that mode must be rejected since
$Q \ll 0.001$ for all values of $l$ considered in our simulations. At the
same time, the figure also clearly illustrates that the solution must be
either $l= 0$ or $l = 1$. We note that if the reported amplitude
uncertainties are multiplied by a factor of 3.5 (to mimic the
possibility that these uncertaities have been perhaps underestimated),
then the solutions $l = 0$ and $l = 1$ become formally acceptable, while
the others can be safely discarded. However, we have no particularly
good reason to believe that the uncertainties on the amplitudes would
have been underestimated by such a relatively large amount in this data
set. Let us consider instead the data at face value and examine if we
can rise up to the challenge offered by this very high S/N observation.

The first obvious possibility to improve the match between the data
points and one ($l = 0$) or the other ($l = 1$) of the possible solutions
is to vary the model parameters within reasonable ranges, in particular
the values of the atmospheric parameters $T_{\rm eff}$ and log $g$ that
we assumed for BAL 090100001. The values of these parameters, as derived
by Green et al. (2006), are $T_{\rm eff}$ = 29,810 $\pm$ 400 K and 
log $g$ = 5.58 $\pm$ 0.05. Explicit calculations within these bounds indicate
that the situation is not changed at the qualitative level: the values of
$\chi^2$ for the $l = 0$ and $l =1$ solutions remain within a factor of
2 of each other as in Table 8, and the values of $Q$ are not
significantly improved.  

The same is true by redoing the modeling exercise using, this time, the
different set of atmospheric parameters derived by Oreiro et al. (2004),
i.e., $T_{\rm eff}$ = 29,450 $\pm$ 500 K and log $g$ = 5.33 $\pm$
0.10. In that case, the solution must again be either $l = 0$ or $l =
1$, and the respective $\chi^2$ values are $1.39 \times 10^2$ and $1.52
\times 10^2$, both comparable to each other, but still way too high to be
formally acceptable as their associated $Q$ values are much smaller than
the criterion level of 0.001. We also experimented with changing the
extinction curve from that of KPNO to one appropriate at sea level and
found very little changes in our theoretical amplitudes. Likewise,
assuming a specific CCD response, instead of using a flat gray response
as we did in our initial calculations, also led to very little
qualitative changes compared to the situation depicted in Figure 12 and
Table 8. (In that latter experiment, we did not use the response curve of
the CCD employed by B05 since it was not available to us, but that of
CCD21 of Steward Observatory as a surrogate.) Thus, we
are unable to find a formally acceptable ($Q > 0.001$) $l$ model for the
main pulsation mode in BAL 090100001, but we still conclude that it must
be either a $l =0$ or a $l = 1$ oscillation.

This failure to model properly the main oscillation observed in BAL
090100001 is somewhat bothersome (assuming again that the reported amplitude
uncertainties are realistic estimates). We remind the reader that
Randall et al. (2005), using the exact same tools as those used in this
paper, have been able to model successfully the 182.42 s mode {\sl
ULTRACAM} data reported by Jeffery et al. (2004) for KPD 2109+4401. We
can think of a potentially important difference between the two cases,
however, and it is the fact that the amplitude of the dominant mode in
BAL 090100001 is so large that nonlinear effects are obviously present
in the form of the first and second harmonics. No such nonlinear
features have been observed in KPD 2109+4401. The theory used to
determine amplitude ratios between different bandpasses is strictly
linear, and it is possible, although this remains unproven, that it
becomes inadequate to treat very large amplitude oscillations such as
the 356.19 s mode in BAL 090100001.

Figures 13, 14, and 15 show our model fits for the triplet of modes
which has been interpreted as a rotationally-split $l = 1$ mode by
B05. Taking again the data at face value, we find (see Table 8) that the
354.20 s mode has $l = 2$, the 354.01 s mode has a value of $l$ of
either 1 or 2, and the 353.81 s component has a value of $l$ of either
1, 2, or 4. If, however, the amplitude uncertainties are somewhat
underestimated, then the $l = 1$ possibility for the 354.20 s mode
should not be dismissed too hastily because its $Q$ value would not be 
much smaller than the passage value of $Q = 0.001$. Because of this, we
prefer to be conservative and conclude that if the observed triplet of
modes is indeed due to rotational splitting, then it must have a value
of either $l = 1$ or $l = 2$. In the latter case, an unfavorable
observation angle or some other cause could perhaps hide two components
of the quintuplet. 

We have thus been unable to confirm the suggestion of B05 that the
dominant 356.19 s mode in BAL 090100001 is a radial mode, and that the
354.20, 354.01, and 353.81 s modes are the components of a
rotationally-split $l = 1$ mode. Our results are nevertheless consistent
with that suggestion, but we found that the dominant mode could also
have a degree index of $l =1$ while the triplet could be three
components of a split $l = 2$ mode.

\subsection{PG 0014+067 (J05)}

The case of PG 0014+067 is particularly interesting because a specific
asteroseismic model has been proposed by Brassard et al. (2001)
including, of course, complete mode identification. This model has been
refined subsequently by Charpinet et al. (2005c) who exploited a higher
quality (white light) data set and found the same mode identification. 
In both cases, in the search in parameter space for the optimal model, it was
explicitly assumed that the observed modes had to belong to degree
indices from $l = 0$ up to and including $l = 3$. The inclusion of $l = 3$ 
was deemed necessary because the observed mode density in PG 0014+067
is too large to be explained solely in terms of modes with $l = 0$, 1,
and 2. In other words, there are more observed modes in the relevant period
window of PG 0014+067 than there are theoretical periods with $l$ = 0, 1,
and 2, so there seemed to be no choice but conclude that some additional modes
with $l \geq 3$ are excited to visible levels in that star. Except for
the restriction to $l$ values from 0 to 3, no a priori constraints were
imposed in the asteroseismological exercises carried out by Brassard et
al. (2001) and Charpinet et al. (2005c), and the actual mode
identification (the determination of both the radial order $k$ and the
degree index $l$ for each observed pulsation) came out as a natural
byproduct of the method. 

The results of Brassard et al. (2001) correspond to the first claimed
successful asteroseismological exercise for a pulsating sdB star, and
the importance of testing the proposed seismic model of PG 0014+067
cannot be overstated. It is in this spirit that J05 decided to attempt
testing the $l$ identification inferred in Brassard et al. (2001) using
multicolor photometry. Even though the authors of J05 realized fully
that, at $V \simeq$ 16, PG 0014+067 was going to be a challenge, they
thought the matter to be sufficiently important to invest almost 30 h of
WHT time using the {\sl ULTRACAM} camera. It would have been difficult
to do better than they did on this from an observational point of
view. As in the previous targets studied in this paper, the details of
their data set are given in Table 1. J05 isolated some 10 distinct
pulsation modes in PG 0014+067, and we modeled each one of these
pulsations using the parameters listed in Table 2. 

Not unexpectedly, our analysis reveals that most of the observed modes
in PG 0014+067 cannot be tested for $l$ identification as the measured
amplitudes usually do not have high enough S/N ratios. The details of
our quantitative analysis are presented in Table 9 for the 10 modes of
interest. At the same time, our work also indicates that partial $l$
discrimination is possible for the three largest amplitude modes
uncovered by J05, and we show our model fits for these modes in Figure
16 (141.01 s), Figure 17 (141.06 s), and Figure 18 (146.50 s). We find
that the observed amplitudes of the 141.01 s mode can be accommodated if
the degree index $l$ has a value of either 0, 1, 2, or 4. This agrees
with the conclusion of J05 that this mode (the $f_{12}$ mode in their
notation) cannot have a value of $l = 3$, and indeed we find that $Q =
1.2 \times 10^{-21}$ for that model, much too small to be acceptable. We
also find that the 141.06 s mode ($f_{11}$ in J05) has a value of $l$ of
either 0, 1, or 2, again in agreement with the inference made in
J05. Finally, we find that the observed amplitude-wavelength
distribution of the 146.50 s mode ($f_{9}$ in J05) can be quantitatively
explained if the mode has a $l$ value of either 0, 1, or 2. This again
is consistent with J05 who found that this mode cannot be a mode with
$l = 3$ or $l = 4$. 

The most interesting result of this analysis is the conclusion, first
put forward by J05, that the 141.06 s mode (observed as the 141.07 s
mode in the lower resolution white light data of Brassard et al. 2001)
cannot have a degree index of $l = 3$. Given that Brassard et al. (2001)
formally identified this mode as a $l = 3$ pulsation, this implies that
a reevalution of their seismic model is warranted. It should be pointed
out that at the time of the analysis of Brassard et al. (2001), the
theory of multicolor photometry had not yet been applied to realistic
models of pulsating sdB stars. Using an Eddington limb darkening law
(instead of an exact form coming from detailed sdB model atmospheres),
Brassard et al. (2001) estimated that the visibility factor would be
equal to 1.0000, 0.7083, 0.3250, 0.0625, and 0.0208 respectively for
modes with $l = 0$, 1, 2, 3, and 4. Hence, the $l = 3$ modes would be
more than three times ``more visible'' than their $l = 4$ counterparts,
and it was deemed natural to limit the search to modes with $l = 3$
beyond those with $l$ = 0, 1, and 2 since this was now sufficient to
account for the relatively high density of observed modes.

What we learned in Randall et al. (2005), among many other things, is that
the visibility of a $p$-mode with $l = 4$ is, in fact, substantially
larger than that of a $l = 3$ mode in a sdB pulsator {\it in the optical
domain}. The discussion of Figure 1 in Randall et al. (2005) is quite
explicit about this. Hence, it will be necessary in future
asteroseismological exercises concerning EC 14026 pulsators (especially
those showing more modes than could be accommodated by invoking the
presence of only modes with $l = 0$, 1 , and 2) to include the
possibility that the detected modes could also have values of $l = 4$ as 
they have a higher probability of being detected than the $l = 3$
modes. In particular, a reanalysis of the data presented by Brassard et
al. (2001) and Charpinet et al. (2005c) on PG 0014+067 including this 
possibility must be carried out. This is being done and will be reported in
due time.   

\section{DISCUSSION}

We have presented in this paper a quantitative and homogeneous analysis
of the broadband multicolor photometric data sets gathered so far on
rapidly pulsating sdB (EC 14026) stars. This analysis was carried out
within the theoretical framework developed by Randall et al. (2005) with
the goal of partial mode identification, i.e., the determination of the
degree index $l$ of each of the observed pulsation modes. With the
exception of the very high S/N ratio $u'g'r'$/{\sl ULTRACAM} data set
obtained by Jeffery et al. (2004) on KPD 2109+4401 (an analysis of which
will be presented in a separate publication), we considered all
available data that we could find in the literature. This consists of
7 distinct data sets pertaining to 6 different EC 14016 stars and
involving 41 pulsation modes.

Our final results are summarized in Table 10 where we list the name of
the target star (1st column), the period of the mode of interest (2nd
column), the acceptable values of $l$ for that mode (3rd column), and a
qualifier as to the level of mode discrimination achieved (4th column).   
It can be seen that no completely convincing and unambiguous $l$
identification has been possible on the basis of the available data,
although we came close to that goal in the case of two modes in KPD
2109+4401 (K98) and three modes in BAL 090100001 (B05) for which only
two possible values of $l$ were found to be consistent with the observed
amplitude-wavelength distributions. At the same time, we found that no
mode discrimination (between values from $l =0$ to $l =5$) was possible
for 16 out of 41 modes. There is no doubt that these disappointing (but
perhaps not totally unexpected) results are due to the fact that the
available data sets were of insufficient sensitivity. The exception 
is the data set of B05 which features a remarkably high S/N value for
the largest amplitude mode (356.19 s) uncovered in that study. We
speculate here that nonlinear effects (observed in the form of the first
and second harmonic of that mode) may have undermined our ability to
model properly the amplitude-wavelength behavior of that particular
mode. The latter shows an exceptionally large amplitude by EC 14026 star
standards. We recall, in this context, that the feasibility of our
approach has been demonstrated beyond any doubt by Randall et al. (2005)
in their analysis of the 182.42 s mode measured by Jeffery et al. (2004)
in KPD 2109+4401. Hence, the case of the 356.19 s mode in BAL 090100001
remains enigmatic.

It is interesting to examine the statistics of the numbers shown in
Table 10. Considering only those cases where partial mode discrimination
has been possible, we find that the $l= 0$ solution comes up 19 times,
the $l = 1$ solution 23 times, the $l = 2$ solution 24 times, the $l =3$
solution 3 times, the $l =4$ solution 13 times, and the $l =5$ solution
once. While no single $l$ index identification has been achieved in our
study, one can argue, from a strictly statistical point of view, that
these results are certainly consistent with the view that the pulsations
seen in rapidly pulsating sdB stars are low-degree modes, mostly with
values of $l =0$, 1, and 2. This is not a great surprise by any means,
but it confirms the early interpretation of the EC 14026 phenomenon in
terms of low-order, low-degree $p$-mode pulsations as presented by
Charpinet et al. (1997). Moreover, the higher frequency of $l =4$
solutions compared to that of $l = 3$ solutions is consistent with
the finding of Randall et al. (2005) that the former modes are more
visible than the latter (in the optical domain).

In conclusion, we find that in order to exploit the full potential of
multicolor photometry for EC 14026 pulsators, it will be necessary in
the future to gather higher S/N ratio observations than has been done up
to now. This is certainly within reach, however, as our results clearly
imply. Furthermore, it will be necessary in future asteroseismological
exercises such as those carried out by Brassard et al. (2001) and
Charpinet et al. (2005c) to include, as required, theoretical modes with
$l = 4$ since those have a higher visibility factor in sdB stars than $l =
3$ modes.

\acknowledgements

This work was supported in part by the NSERC of Canada. G. Fontaine also
acknowledges the contribution of the Canada Research Chair Program. 
P. Bergeron is a Cottrell Scholar of the Research Corporation.     

\clearpage

\begin{deluxetable}{lccccccc}
\rotate
\tablewidth{0pt}
\tablecaption{Basic characteristics of the data sets used in this paper}
\tablehead{
\colhead{Object} &
\colhead{$V$} &
\colhead{Hours} &
\colhead{Filters} &
\colhead{Instrument} &
\colhead{Telescope} &
\colhead{Location} &
\colhead{Reference}
}
 
\startdata
KPD 2109+4401 & 13.38 & 26.6 & $UBVR$ & 4-channel Steining & Struve &
Mt. Locke & K98\\
... & ... & ... & ... & photometer & 2.1 m & Texas & ...\\
PG 1618+563B & 13.52 & 1.5 & $UBV$ & 3-channel Tromso & NOT & La Palma &
S00\\
... & ... & ... & ... & photometer & 2.5 m & Spain & ...\\
PG 1605+072 & 12.92 & 12.3 & $UV,B,R,NIR$ & 4-channel {\sl BUSCA} & 2.2 m &
Calar Alto & F03\\
... & ... & ... & ... & photometer & ... & Spain & ...\\
HS 0039+4302 & 15.5 & 16.2 & $u'g'r'$ & {\sl ULTRACAM} & WHT & La Palma
& J04\\
... & ... & ... & ... & ... & 4.2 m & Spain & ...\\
BAL 090100001 & 11.8($B$) & 9.3 & $uvby$ & Str\"omgren & 0.9 m & Sierra
Nevada & O05\\
... & ... & ... & ... & photometer & ... & Spain & ...\\ 
BAL 090100001 & 11.8($B$) & 126 & $UBVR$ & CCD & 0.6 m & Mt. Suhora & B05\\
... & ... & ... & ... & ... & ... & Poland & ...\\ 
PG 0014+067 & 15.9 & 29.5 & $u'g'r'$ & {\sl ULTRACAM} & WHT & La Palma
& J05\\
... & ... & ... & ... & ... & 4.2 m & Spain & ...\\
\enddata
\end{deluxetable}
\clearpage

\begin{deluxetable}{lccccc}
\tablecolumns{6}
\tablewidth{0pt}
\tablecaption{Atmospheric parameters used in the computations of the
  theoretical amplitudes}
\tablehead{
\colhead{Object} &
\colhead{$T_{\rm eff}$ (K)} &
\colhead{log $g$} &
\colhead{Filters} &
\colhead{Instrument response} &
\colhead{Extinction}
}
 
\startdata
KPD 2109+4401 & 31380 & 5.65 & $UBVR$ & gray & KPNO\\
...           & ...  & ...   & (Steining) & ... & ...\\ 
PG 1618+563B &  34320 & 5.79 & $UBV$ & gray & KPNO\\
...           & ...  & ...   & (Johnson) & ... & ...\\ 
PG 1605+072 &   32520 & 5.27 & $UV,B,R,NIR$ & {\sl BUSCA} & KPNO\\
...           & ...  & ...   & ...  & (Cordes) & ...\\ 
HS 0039+4302 &  32320 & 5.68 & $u'g'r'$ & {\sl ULTRACAM} & KPNO\\
...           & ...  & ...   & (Sloan)  & (Dhillon) & ...\\ 
BAL 090100001 & 29810 & 5.58 & $uvby$ & gray & KPNO\\
...           & ...  & ...   & (Str\"omgren)  & ... & ...\\ 
BAL 090100001 & 29810 & 5.58 & $UBVR$ & gray & KPNO\\
...           & ...  & ...   & (Johnson/Cousins)  & ... & ...\\ 
PG 0014+067 &   34130 & 5.77 & $u'g'r'$ & {\sl ULTRACAM} & KPNO\\
...           & ...  & ...   & (Sloan)  & (Dhillon) & ...\\ 
\enddata
\end{deluxetable}
\clearpage

\begin{deluxetable}{cccccccc}
\tablecolumns{8}
\tablewidth{0pt}
\tablecaption{Fits of predicted $UBVR$ amplitudes to those observed for
  the pulsation modes of KPD 2109+4401 reported by K98}
\tablehead{
\colhead{Period} &
\colhead{$U$} &
\colhead{$B$} &
\colhead{$V$} &
\colhead{$R$} &
\colhead{$l$} &
\colhead{$\chi^2$} &
\colhead{$Q$}\\
\colhead{(s)} &
\colhead{(mmag)} &
\colhead{(mmag)} &
\colhead{(mmag)} &
\colhead{(mmag)} &
\colhead{ } &
\colhead{ } &
\colhead{ }
}
 
\startdata
182.42 & 5.300$\pm$0.105 & 4.200$\pm$0.105 & 3.900$\pm$0.085 & 3.700$\pm$0.155 & ... & ... & ...\\
 ...   & 5.444 & 4.091 & 3.836 & 3.722 &  0 &  3.54e+00 &  3.15e$-$01\\
 ...   & 5.352 & 4.116 & 3.888 & 3.785 &  1 &  1.20e+00 &  7.53e$-$01\\
 ...   & 5.242 & 4.143 & 3.947 & 3.846 &  2 &  1.80e+00 &  6.14e$-$01\\
 ...   & 6.070 & 3.846 & 3.463 & 2.974 &  3 &  1.14e+02 &  1.91e$-$24\\
 ...   & 4.726 & 4.251 & 4.159 & 4.147 &  4 &  4.77e+01 &  2.44e$-$10\\
 ...   & 3.950 & 4.134 & 4.371 & 4.820 &  5 &  2.49e+02 &  1.28e$-$53\\
   &       &       &       &       &   &           &          \\
184.72 & 3.000$\pm$0.105 & 2.600$\pm$0.105 & 2.400$\pm$0.085 & 2.200$\pm$0.155 & ... & ... & ...\\
 ...   & 3.242 & 2.436 & 2.283 & 2.215 &  0 &  9.66e+00 &  2.17e$-$02\\
 ...   & 3.188 & 2.452 & 2.316 & 2.254 &  1 &  6.29e+00 &  9.82e$-$02\\
 ...   & 3.124 & 2.471 & 2.354 & 2.294 &  2 &  3.57e+00 &  3.12e$-$01\\
 ...   & 3.596 & 2.285 & 2.057 & 1.766 &  3 &  6.53e+01 &  4.31e$-$14\\
 ...   & 2.822 & 2.541 & 2.486 & 2.480 &  4 &  7.49e+00 &  5.79e$-$02\\
 ...   & 2.366 & 2.477 & 2.622 & 2.892 &  5 &  6.46e+01 &  6.14e$-$14\\
   &       &       &       &       &   &           &          \\
184.75 & 3.900$\pm$0.105 & 3.400$\pm$0.105 & 3.300$\pm$0.085 & 3.100$\pm$0.155 & ... & ... & ...\\
 ...   & 4.328 & 3.252 & 3.048 & 2.958 &  0 &  2.83e+01 &  3.21e$-$06\\
 ...   & 4.256 & 3.274 & 3.092 & 3.010 &  1 &  1.93e+01 &  2.38e$-$04\\
 ...   & 4.172 & 3.300 & 3.144 & 3.063 &  2 &  1.11e+01 &  1.14e$-$02\\
 ...   & 4.786 & 3.042 & 2.738 & 2.350 &  3 &  1.50e+02 &  2.71e$-$32\\
 ...   & 3.776 & 3.400 & 3.327 & 3.318 &  4 &  3.48e+00 &  3.23e$-$01\\
 ...   & 3.174 & 3.323 & 3.517 & 3.880 &  5 &  8.02e+01 &  2.77e$-$17\\
   &       &       &       &       &   &           &          \\
191.85 & 2.700$\pm$0.105 & 2.300$\pm$0.105 & 2.400$\pm$0.085 & 2.200$\pm$0.155 & ... & ... & ...\\
 ...   & 3.038 & 2.280 & 2.136 & 2.072 &  0 &  2.07e+01 &  1.19e$-$04\\
 ...   & 2.986 & 2.298 & 2.169 & 2.112 &  1 &  1.51e+01 &  1.70e$-$03\\
 ...   & 2.924 & 2.317 & 2.207 & 2.151 &  2 &  9.81e+00 &  2.03e$-$02\\
 ...   & 3.342 & 2.144 & 1.927 & 1.653 &  3 &  8.30e+01 &  7.08e$-$18\\
 ...   & 2.646 & 2.389 & 2.340 & 2.335 &  4 &  2.25e+00 &  5.22e$-$01\\
 ...   & 2.228 & 2.336 & 2.478 & 2.739 &  5 &  3.33e+01 &  2.80e$-$07\\
   &       &       &       &       &   &           &          \\
196.31 & 7.200$\pm$0.105 & 5.600$\pm$0.105 & 5.200$\pm$0.085 & 4.800$\pm$0.155 & ... & ... & ...\\
 ...   & 7.306 & 5.481 & 5.131 & 4.978 &  0 &  4.29e+00 &  2.31e$-$01\\
 ...   & 7.170 & 5.518 & 5.207 & 5.070 &  1 &  3.74e+00 &  2.91e$-$01\\
 ...   & 7.004 & 5.557 & 5.294 & 5.160 &  2 &  1.03e+01 &  1.65e$-$02\\
 ...   & 8.084 & 5.215 & 4.685 & 4.016 &  3 &  1.47e+02 &  1.38e$-$31\\
 ...   & 6.294 & 5.693 & 5.577 & 5.569 &  4 &  1.20e+02 &  9.75e$-$26\\
 ...   & 5.246 & 5.505 & 5.849 & 6.473 &  5 &  5.22e+02 &  8.49$-$113\\
   &       &       &       &       &   &           &          \\
196.69 & 0.800$\pm$0.105 & 0.700$\pm$0.105 & 0.700$\pm$0.085 & 0.800$\pm$0.155 & ... & ... & ...\\
 ...   & 0.920 & 0.690 & 0.646 & 0.627 &  0 &  2.97e+00 &  3.97e$-$01\\
 ...   & 0.904 & 0.696 & 0.657 & 0.639 &  1 &  2.32e+00 &  5.09e$-$01\\
 ...   & 0.884 & 0.701 & 0.668 & 0.651 &  2 &  1.70e+00 &  6.37e$-$01\\
 ...   & 1.006 & 0.649 & 0.583 & 0.500 &  3 &  9.72e+00 &  2.11e$-$02\\
 ...   & 0.802 & 0.726 & 0.711 & 0.710 &  4 &  4.15e$-$01 &  9.37e$-$01\\
 ...   & 0.676 & 0.709 & 0.754 & 0.834 &  5 &  1.85e+00 &  6.03e$-$01\\
   &       &       &       &       &   &           &          \\
198.19 & 4.800$\pm$0.105 & 4.000$\pm$0.105 & 4.100$\pm$0.085 & 4.000$\pm$0.155 & ... & ... & ...\\
 ...   & 5.318 & 3.989 & 3.733 & 3.622 &  0 &  4.90e+01 &  1.33e$-$10\\
 ...   & 5.224 & 4.021 & 3.794 & 3.694 &  1 &  3.32e+01 &  2.91e$-$07\\
 ...   & 5.108 & 4.055 & 3.863 & 3.765 &  2 &  1.90e+01 &  2.77e$-$04\\
 ...   & 5.832 & 3.771 & 3.387 & 2.902 &  3 &  2.22e+02 &  7.82e$-$48\\
 ...   & 4.614 & 4.177 & 4.092 & 4.087 &  4 &  6.29e+00 &  9.85e$-$02\\
 ...   & 3.880 & 4.073 & 4.330 & 4.795 &  5 &  1.11e+02 &  7.10e$-$24\\

\enddata
\end{deluxetable}
\clearpage

\begin{deluxetable}{ccccccc}
\tablecolumns{7}
\tablewidth{0pt}
\tablecaption{Fits of predicted $UBV$ amplitudes to those observed for
  the pulsation modes of PG 1618+563B reported by S00}
\tablehead{
\colhead{Period} &
\colhead{$U$} &
\colhead{$B$} &
\colhead{$V$} &
\colhead{$l$} &
\colhead{$\chi^2$} &
\colhead{$Q$}\\
\colhead{(s)} &
\colhead{(mmag)} &
\colhead{(mmag)} &
\colhead{(mmag)} &
\colhead{ } &
\colhead{ } &
\colhead{ }
}
 
\startdata
139.30 & 1.500$\pm$0.500 & 1.200$\pm$0.600 & 0.600$\pm$0.500 & ... & ... & ...\\
 ...   & 1.292 & 1.050 & 0.983 &  0 &  8.23e$-$01 &  6.63e$-$01\\
 ...   & 1.276 & 1.056 & 0.995 &  1 &  8.82e$-$01 &  6.43e$-$01\\
 ...   & 1.254 & 1.062 & 1.008 &  2 &  9.61e$-$01 &  6.19e$-$01\\
 ...   & 1.442 & 1.007 & 0.853 &  3 &  3.74e$-$01 &  8.30e$-$01\\
 ...   & 1.168 & 1.079 & 1.055 &  4 &  1.31e+00 &  5.19e$-$01\\
 ...   & 1.078 & 1.063 & 1.105 &  5 &  1.79e+00 &  4.09e$-$01\\
  &       &       &       &   &           &          \\
143.90 & 4.600$\pm$0.700 & 4.300$\pm$0.800 & 2.600$\pm$1.000 & ... & ... & ...\\
 ...   & 4.608 & 3.745 & 3.504 &  0 &  1.30e+00 &  5.23e$-$01\\
 ...   & 4.566 & 3.778 & 3.559 &  1 &  1.35e+00 &  5.10e$-$01\\
 ...   & 4.508 & 3.820 & 3.627 &  2 &  1.43e+00 &  4.89e$-$01\\
 ...   & 4.940 & 3.478 & 2.949 &  3 &  1.41e+00 &  4.93e$-$01\\
 ...   & 4.278 & 3.958 & 3.872 &  4 &  2.01e+00 &  3.66e$-$01\\
 ...   & 4.048 & 3.994 & 4.159 &  5 &  3.20e+00 &  2.02e$-$01\\
\enddata
\end{deluxetable}
\clearpage

\begin{deluxetable}{cccccccc}
\tablecolumns{8}
\tablewidth{0pt}
\tablecaption{Fits of predicted {\sl BUSCA} amplitudes to those observed for
  the pulsation modes of PG 1605+072 reported by F03}
\tablehead{
\colhead{Period} &
\colhead{$UV$} &
\colhead{$B$} &
\colhead{$R$} &
\colhead{$NIR$} &
\colhead{$l$} &
\colhead{$\chi^2$} &
\colhead{$Q$}\\
\colhead{(s)} &
\colhead{(mmag)} &
\colhead{(mmag)} &
\colhead{(mmag)} &
\colhead{(mmag)} &
\colhead{ } &
\colhead{ } &
\colhead{ }
}
 
\startdata
440.51 &  8.92$\pm$ 1.22 &  6.85$\pm$ 1.22 &  6.85$\pm$ 0.90 &  6.66$\pm$ 1.10 & ... & ... & ...\\
 ...   &  8.80 &  7.15 &  6.99 &  6.29 &  0 &  2.10e$-$01 &  9.76e$-$01\\
 ...   &  8.62 &  7.18 &  7.03 &  6.41 &  1 &  2.28e$-$01 &  9.73e$-$01\\
 ...   &  8.38 &  7.22 &  7.08 &  6.51 &  2 &  3.72e$-$01 &  9.46e$-$01\\
 ...   & 10.21 &  7.55 &  6.44 &  3.11 &  3 &  1.21e+01 &  6.97e$-$03\\
 ...   &  7.55 &  7.15 &  7.16 &  7.12 &  4 &  1.62e+00 &  6.55e$-$01\\
 ...   &  6.04 &  6.40 &  6.99 &  8.44 &  5 &  8.40e+00 &  3.84e$-$02\\
   &       &       &       &       &   &           &          \\
475.82 & 19.76$\pm$ 1.22 & 16.94$\pm$ 1.22 & 16.68$\pm$ 0.90 & 17.18$\pm$ 1.10 & ... & ... & ...\\
 ...   & 21.21 & 17.19 & 16.80 & 15.10 &  0 &  5.09e+00 &  1.66e$-$01\\
 ...   & 20.73 & 17.27 & 16.92 & 15.42 &  1 &  3.36e+00 &  3.40e$-$01\\
 ...   & 20.17 & 17.41 & 17.07 & 15.70 &  2 &  2.26e+00 &  5.19e$-$01\\
 ...   & 24.16 & 18.05 & 15.40 &  7.58 &  3 &  9.26e+01 &  5.92e$-$20\\
 ...   & 18.20 & 17.28 & 17.32 & 17.27 &  4 &  2.26e+00 &  5.20e$-$01\\
 ...   & 14.58 & 15.52 & 17.00 & 20.62 &  5 &  2.95e+01 &  1.79e$-$06\\
   &       &       &       &       &   &           &          \\
481.75 & 36.88$\pm$ 1.22 & 29.27$\pm$ 1.22 & 28.45$\pm$ 0.90 & 28.72$\pm$ 1.10 & ... & ... & ...\\
 ...   & 37.05 & 30.03 & 29.33 & 26.36 &  0 &  6.00e+00 &  1.12e$-$01\\
 ...   & 36.20 & 30.14 & 29.53 & 26.92 &  1 &  4.98e+00 &  1.74e$-$01\\
 ...   & 35.16 & 30.36 & 29.77 & 27.40 &  2 &  6.44e+00 &  9.22e$-$02\\
 ...   & 42.50 & 31.80 & 27.13 & 13.40 &  3 &  2.23e+02 &  4.30e$-$48\\
 ...   & 31.62 & 30.06 & 30.13 & 30.06 &  4 &  2.41e+01 &  2.34e$-$05\\
 ...   & 25.18 & 26.83 & 29.41 & 35.70 &  5 &  1.38e+02 &  9.56e$-$30\\
   &       &       &       &       &   &           &          \\
503.70 & 22.59$\pm$ 1.22 & 22.45$\pm$ 1.22 & 18.79$\pm$ 0.90 & 18.02$\pm$ 1.10 & ... & ... & ...\\
 ...   & 24.48 & 19.80 & 19.34 & 17.36 &  0 &  7.83e+00 &  4.96e$-$02\\
 ...   & 23.89 & 19.89 & 19.48 & 17.76 &  1 &  6.18e+00 &  1.03e$-$01\\
 ...   & 23.18 & 20.05 & 19.67 & 18.11 &  2 &  5.04e+00 &  1.69e$-$01\\
 ...   & 28.05 & 21.12 & 18.03 &  9.00 &  3 &  8.97e+01 &  2.53e$-$19\\
 ...   & 20.84 & 19.86 & 19.92 & 19.90 &  4 &  1.11e+01 &  1.12e$-$02\\
 ...   & 16.52 & 17.65 & 19.39 & 23.61 &  5 &  6.68e+01 &  2.09e$-$14\\
   &       &       &       &       &   &           &          \\
528.70 &  7.65$\pm$ 1.22 &  8.25$\pm$ 1.22 &  6.79$\pm$ 0.90 &  6.88$\pm$ 1.10 & ... & ... & ...\\
 ...   &  8.84 &  7.13 &  6.96 &  6.24 &  0 &  2.16e+00 &  5.39e$-$01\\
 ...   &  8.62 &  7.17 &  7.03 &  6.41 &  1 &  1.66e+00 &  6.45e$-$01\\
 ...   &  8.35 &  7.24 &  7.10 &  6.55 &  2 &  1.22e+00 &  7.48e$-$01\\
 ...   &  9.99 &  7.58 &  6.47 &  3.27 &  3 &  1.50e+01 &  1.83e$-$03\\
 ...   &  7.51 &  7.19 &  7.21 &  7.22 &  4 &  1.09e+00 &  7.80e$-$01\\
 ...   &  5.96 &  6.40 &  7.05 &  8.61 &  5 &  6.79e+00 &  7.91e$-$02\\
\enddata
\end{deluxetable}
\clearpage

\begin{deluxetable}{ccccccc}
\tablecolumns{7}
\tablewidth{0pt}
\tablecaption{Fits of predicted {\sl ULTRACAM} amplitudes to those observed for
  the pulsation modes of HS 0039+4302 reported by J04}
\tablehead{
\colhead{Period} &
\colhead{$u'$} &
\colhead{$g'$} &
\colhead{$r'$} &
\colhead{$l$} &
\colhead{$\chi^2$} &
\colhead{$Q$}\\
\colhead{(s)} &
\colhead{(mmag)} &
\colhead{(mmag)} &
\colhead{(mmag)} &
\colhead{ } &
\colhead{ } &
\colhead{ }
}
 
\startdata

134.44 & 0.890$\pm$0.156 & 0.730$\pm$0.063 & 0.600$\pm$0.081 & ... & ... & ...\\
 ...   & 0.914 & 0.690 & 0.660 &  0 &  9.81e$-$01 &  6.12e$-$01\\
 ...   & 0.900 & 0.690 & 0.663 &  1 &  9.99e$-$01 &  6.07e$-$01\\
 ...   & 0.886 & 0.693 & 0.666 &  2 &  1.02e+00 &  6.02e$-$01\\
 ...   & 1.178 & 0.676 & 0.545 &  3 &  4.62e+00 &  9.95e$-$02\\
 ...   & 0.794 & 0.697 & 0.687 &  4 &  1.81e+00 &  4.04e$-$01\\
 ...   & 0.678 & 0.676 & 0.735 &  5 &  5.38e+00 &  6.80e$-$02\\
  &       &       &       &   &           &          \\
180.13 & 1.590$\pm$0.156 & 1.350$\pm$0.063 & 1.410$\pm$0.081 & ... & ... & ...\\
 ...   & 1.820 & 1.363 & 1.300 &  0 &  4.06e+00 &  1.31e$-$01\\
 ...   & 1.780 & 1.367 & 1.311 &  1 &  3.06e+00 &  2.16e$-$01\\
 ...   & 1.732 & 1.374 & 1.321 &  2 &  2.18e+00 &  3.36e$-$01\\
 ...   & 2.152 & 1.358 & 1.093 &  3 &  2.83e+01 &  7.12e$-$07\\
 ...   & 1.540 & 1.383 & 1.372 &  4 &  6.04e$-$01 &  7.39e$-$01\\
 ...   & 1.322 & 1.340 & 1.487 &  5 &  3.89e+00 &  1.43e$-$01\\
  &       &       &       &   &           &          \\
181.89 & 3.590$\pm$0.156 & 2.960$\pm$0.063 & 2.640$\pm$0.081 & ... & ... & ...\\
 ...   & 3.820 & 2.859 & 2.728 &  0 &  5.90e+00 &  5.23e$-$02\\
 ...   & 3.734 & 2.868 & 2.749 &  1 &  4.82e+00 &  8.98e$-$02\\
 ...   & 3.628 & 2.878 & 2.768 &  2 &  4.22e+00 &  1.21e$-$01\\
 ...   & 4.542 & 2.874 & 2.313 &  3 &  5.54e+01 &  9.15e$-$13\\
 ...   & 3.216 & 2.890 & 2.867 &  4 &  1.48e+01 &  6.10e$-$04\\
 ...   & 2.750 & 2.788 & 3.097 &  5 &  6.83e+01 &  1.51e$-$15\\
  &       &       &       &   &           &          \\
182.79 & 2.390$\pm$0.156 & 2.000$\pm$0.063 & 1.800$\pm$0.081 & ... & ... & ...\\
 ...   & 2.582 & 1.932 & 1.843 &  0 &  2.95e+00 &  2.28e$-$01\\
 ...   & 2.524 & 1.938 & 1.858 &  1 &  2.21e+00 &  3.31e$-$01\\
 ...   & 2.452 & 1.946 & 1.871 &  2 &  1.67e+00 &  4.35e$-$01\\
 ...   & 3.064 & 1.941 & 1.562 &  3 &  2.82e+01 &  7.66e$-$07\\
 ...   & 2.174 & 1.954 & 1.939 &  4 &  5.37e+00 &  6.82e$-$02\\
 ...   & 1.860 & 1.886 & 2.096 &  5 &  2.81e+01 &  7.78e$-$07\\
  &       &       &       &   &           &          \\
192.58 & 5.730$\pm$0.156 & 4.350$\pm$0.063 & 4.050$\pm$0.081 & ... & ... & ...\\
 ...   & 5.764 & 4.309 & 4.108 &  0 &  9.90e$-$01 &  6.10e$-$01\\
 ...   & 5.624 & 4.320 & 4.140 &  1 &  1.93e+00 &  3.80e$-$01\\
 ...   & 5.454 & 4.337 & 4.170 &  2 &  5.38e+00 &  6.79e$-$02\\
 ...   & 6.776 & 4.351 & 3.501 &  3 &  9.09e+01 &  1.79e$-$20\\
 ...   & 4.824 & 4.352 & 4.321 &  4 &  4.49e+01 &  1.77e$-$10\\
 ...   & 4.122 & 4.191 & 4.671 &  5 &  1.71e+02 &  6.19e$-$38\\
  &       &       &       &   &           &          \\
234.11 & 6.110$\pm$0.156 & 4.930$\pm$0.063 & 4.620$\pm$0.081 & ... & ... & ...\\
 ...   & 6.536 & 4.843 & 4.607 &  0 &  9.38e+00 &  9.17e$-$03\\
 ...   & 6.328 & 4.862 & 4.658 &  1 &  3.32e+00 &  1.90e$-$01\\
 ...   & 6.072 & 4.888 & 4.704 &  2 &  1.59e+00 &  4.53e$-$01\\
 ...   & 7.290 & 4.953 & 3.993 &  3 &  1.17e+02 &  3.37e$-$26\\
 ...   & 5.336 & 4.904 & 4.890 &  4 &  3.59e+01 &  1.60e$-$08\\
 ...   & 4.554 & 4.699 & 5.315 &  5 &  1.87e+02 &  3.07e$-$41\\

\enddata
\end{deluxetable}
\clearpage

\begin{deluxetable}{cccccccc}
\tablecolumns{8}
\tablewidth{0pt}
\tablecaption{Fits of predicted $uvby$ amplitudes to those observed for
  the pulsation modes of BAL 090100001 reported by O05}
\tablehead{
\colhead{Period} &
\colhead{$u$} &
\colhead{$v$} &
\colhead{$b$} &
\colhead{$y$} &
\colhead{$l$} &
\colhead{$\chi^2$} &
\colhead{$Q$}\\
\colhead{(s)} &
\colhead{(mmag)} &
\colhead{(mmag)} &
\colhead{(mmag)} &
\colhead{(mmag)} &
\colhead{ } &
\colhead{ } &
\colhead{ }
}
 
\startdata

351.70 & 16.73$\pm$ 0.08 & 17.43$\pm$ 0.09 & 16.84$\pm$ 0.09 &
   15.76$\pm$ 0.08 &
 ... & ... & ...\\
 ...   & 21.02 & 13.97 & 15.27 & 13.27 &  0 &  5.62e+03 &  0.00e+00\\
 ...   & 20.24 & 14.63 & 15.63 & 14.07 &  1 &  3.51e+03 &  0.00e+00\\
 ...   & 19.37 & 15.30 & 15.93 & 14.83 &  2 &  1.89e+03 &  0.00e+00\\
 ...   & 20.87 & 14.83 & 15.03 & 13.15 &  3 &  4.99e+03 &  0.00e+00\\
 ...   & 17.27 & 16.26 & 16.60 & 16.32 &  4 &  2.71e+02 &  1.90e$-$58\\
 ...   & 14.49 & 16.21 & 16.29 & 18.70 &  5 &  2.35e+03 &  0.00e+00\\
   &       &       &       &       &   &           &          \\
356.30 & 49.36$\pm$ 0.11 & 41.34$\pm$ 0.11 & 38.55$\pm$ 0.10 &
   38.03$\pm$ 0.10 &
 ... & ... & ...\\
 ...   & 54.58 & 36.23 & 39.63 & 34.40 &  0 &  5.84e+03 &  0.00e+00\\
 ...   & 52.05 & 37.63 & 40.20 & 36.19 &  1 &  2.34e+03 &  0.00e+00\\
 ...   & 49.34 & 39.00 & 40.62 & 37.81 &  2 &  8.86e+02 &  8.47$-$192\\
 ...   & 54.02 & 38.46 & 38.99 & 34.10 &  3 &  4.04e+03 &  0.00e+00\\
 ...   & 43.19 & 40.71 & 41.56 & 40.88 &  4 &  4.90e+03 &  0.00e+00\\
 ...   & 35.77 & 40.04 & 40.20 & 46.21 &  5 &  2.24e+04 &  0.00e+00\\
\enddata
\end{deluxetable}
\clearpage

\begin{deluxetable}{cccccccc}
\tablecolumns{8}
\tablewidth{0pt}
\tablecaption{Fits of predicted $UBVR$ amplitudes to those observed for
  the pulsation modes of BAL 090100001 reported by B05}
\tablehead{
\colhead{Period} &
\colhead{$U$} &
\colhead{$B$} &
\colhead{$V$} &
\colhead{$R$} &
\colhead{$l$} &
\colhead{$\chi^2$} &
\colhead{$Q$}\\
\colhead{(s)} &
\colhead{(mmag)} &
\colhead{(mmag)} &
\colhead{(mmag)} &
\colhead{(mmag)} &
\colhead{ } &
\colhead{ } &
\colhead{ }
}
 
\startdata

356.19 & 75.23$\pm$ 0.25 & 57.71$\pm$ 0.18 & 53.34$\pm$ 0.16 & 50.26$\pm$ 0.17 & ... & ... & ...\\
 ...   & 77.92 & 58.14 & 52.05 & 49.37 &  0 &  2.15e+02 &  2.68e$-$46\\
 ...   & 73.18 & 57.96 & 53.38 & 51.31 &  1 &  1.07e+02 &  3.84e$-$23\\
 ...   & 68.50 & 57.76 & 54.53 & 52.85 &  2 &  1.01e+03 &  2.91$-$219\\
 ...   & 77.85 & 59.92 & 52.99 & 46.11 &  3 &  8.61e+02 &  2.11$-$186\\
 ...   & 58.83 & 56.55 & 55.98 & 56.27 &  4 &  5.87e+03 &  0.00e+00\\
 ...   & 46.98 & 49.55 & 55.85 & 63.17 &  5 &  2.08e+04 &  0.00e+00\\
   &       &       &       &       &   &           &          \\
354.20 & 26.71$\pm$ 0.25 & 21.92$\pm$ 0.18 & 20.53$\pm$ 0.16 & 19.92$\pm$ 0.17 & ... & ... & ...\\
 ...   & 29.52 & 22.03 & 19.73 & 18.71 &  0 &  2.02e+02 &  1.57e$-$43\\
 ...   & 27.77 & 22.00 & 20.26 & 19.47 &  1 &  2.81e+01 &  3.55e$-$06\\
 ...   & 26.04 & 21.94 & 20.72 & 20.08 &  2 &  9.53e+00 &  2.30e$-$02\\
 ...   & 29.50 & 22.69 & 20.06 & 17.46 &  3 &  3.61e+02 &  7.16e$-$78\\
 ...   & 22.43 & 21.55 & 21.33 & 21.44 &  4 &  4.03e+02 &  6.13e$-$87\\
 ...   & 17.98 & 18.96 & 21.37 & 24.16 &  5 &  2.14e+03 &  0.00e+00\\
   &       &       &       &       &   &           &          \\
354.01 & 15.82$\pm$ 0.25 & 12.86$\pm$ 0.18 & 11.62$\pm$ 0.16 & 11.57$\pm$ 0.17 & ... & ... & ...\\
 ...   & 17.14 & 12.79 & 11.46 & 10.87 &  0 &  4.61e+01 &  5.27e$-$10\\
 ...   & 16.12 & 12.77 & 11.76 & 11.30 &  1 &  4.91e+00 &  1.78e$-$01\\
 ...   & 15.11 & 12.73 & 12.02 & 11.65 &  2 &  1.51e+01 &  1.71e$-$03\\
 ...   & 17.13 & 13.17 & 11.65 & 10.14 &  3 &  1.01e+02 &  7.47e$-$22\\
 ...   & 13.00 & 12.49 & 12.36 & 12.43 &  4 &  1.78e+02 &  1.91e$-$38\\
 ...   & 10.41 & 10.98 & 12.38 & 13.99 &  5 &  8.02e+02 &  1.76$-$173\\
   &       &       &       &       &   &           &          \\
353.81 &  5.86$\pm$ 0.25 &  5.12$\pm$ 0.18 &  4.71$\pm$ 0.16 &  4.59$\pm$ 0.17 & ... & ... & ...\\
 ...   &  6.75 &  5.04 &  4.51 &  4.28 &  0 &  1.77e+01 &  4.99e$-$04\\
 ...   &  6.35 &  5.03 &  4.63 &  4.45 &  1 &  4.99e+00 &  1.73e$-$01\\
 ...   &  5.96 &  5.02 &  4.74 &  4.59 &  2 &  4.92e$-$01 &  9.21e$-$01\\
 ...   &  6.74 &  5.18 &  4.58 &  3.99 &  3 &  2.57e+01 &  1.13e$-$05\\
 ...   &  5.13 &  4.93 &  4.88 &  4.91 &  4 &  1.42e+01 &  2.69e$-$03\\
 ...   &  4.12 &  4.35 &  4.90 &  5.54 &  5 &  9.93e+01 &  2.23e$-$21\\
   &       &       &       &       &   &           &          \\
350.39 &  2.23$\pm$ 0.25 &  1.81$\pm$ 0.18 &  2.08$\pm$ 0.16 &  1.45$\pm$ 0.17 & ... & ... & ...\\
 ...   &  2.53 &  1.89 &  1.69 &  1.61 &  0 &  8.33e+00 &  3.96e$-$02\\
 ...   &  2.38 &  1.89 &  1.74 &  1.67 &  1 &  6.81e+00 &  7.82e$-$02\\
 ...   &  2.23 &  1.88 &  1.78 &  1.72 &  2 &  6.31e+00 &  9.75e$-$02\\
 ...   &  2.54 &  1.95 &  1.72 &  1.50 &  3 &  7.17e+00 &  6.66e$-$02\\
 ...   &  1.93 &  1.85 &  1.83 &  1.84 &  4 &  9.20e+00 &  2.67e$-$02\\
 ...   &  1.54 &  1.62 &  1.83 &  2.07 &  5 &  2.43e+01 &  2.14e$-$05\\
350.18 &  2.01$\pm$ 0.25 &  1.72$\pm$ 0.18 &  1.31$\pm$ 0.16 &  1.48$\pm$ 0.17 & ... & ... & ...\\
 ...   &  2.15 &  1.60 &  1.44 &  1.36 &  0 &  1.82e+00 &  6.11e$-$01\\
 ...   &  2.02 &  1.60 &  1.47 &  1.41 &  1 &  1.63e+00 &  6.53e$-$01\\
 ...   &  1.89 &  1.59 &  1.50 &  1.46 &  2 &  2.21e+00 &  5.31e$-$01\\
 ...   &  2.15 &  1.65 &  1.46 &  1.27 &  3 &  2.85e+00 &  4.15e$-$01\\
 ...   &  1.62 &  1.56 &  1.54 &  1.55 &  4 &  5.48e+00 &  1.40e$-$01\\
 ...   &  1.30 &  1.37 &  1.54 &  1.74 &  5 &  1.64e+01 &  9.45e$-$04\\
   &       &       &       &       &   &           &          \\
349.83 &  1.94$\pm$ 0.25 &  1.45$\pm$ 0.18 &  1.58$\pm$ 0.16 &  1.45$\pm$ 0.17 & ... & ... & ...\\
 ...   &  2.13 &  1.59 &  1.43 &  1.35 &  0 &  2.45e+00 &  4.84e$-$01\\
 ...   &  2.00 &  1.59 &  1.46 &  1.41 &  1 &  1.26e+00 &  7.38e$-$01\\
 ...   &  1.88 &  1.58 &  1.49 &  1.45 &  2 &  8.91e$-$01 &  8.28e$-$01\\
 ...   &  2.13 &  1.64 &  1.45 &  1.26 &  3 &  3.61e+00 &  3.07e$-$01\\
 ...   &  1.62 &  1.56 &  1.54 &  1.55 &  4 &  2.37e+00 &  5.00e$-$01\\
 ...   &  1.30 &  1.38 &  1.55 &  1.75 &  5 &  9.80e+00 &  2.04e$-$02\\
   &       &       &       &       &   &           &          \\
264.82 &  5.94$\pm$ 0.25 &  4.73$\pm$ 0.18 &  4.16$\pm$ 0.16 &  4.07$\pm$ 0.17 & ... & ... & ...\\
 ...   &  6.09 &  4.64 &  4.21 &  4.01 &  0 &  8.73e$-$01 &  8.32e$-$01\\
 ...   &  5.84 &  4.62 &  4.27 &  4.10 &  1 &  1.00e+00 &  8.01e$-$01\\
 ...   &  5.59 &  4.62 &  4.34 &  4.19 &  2 &  4.15e+00 &  2.46e$-$01\\
 ...   &  6.45 &  4.73 &  4.17 &  3.64 &  3 &  1.04e+01 &  1.55e$-$02\\
 ...   &  4.83 &  4.54 &  4.46 &  4.46 &  4 &  2.96e+01 &  1.70e$-$06\\
 ...   &  3.85 &  4.05 &  4.47 &  4.99 &  5 &  1.17e+02 &  4.05e$-$25\\
   &       &       &       &       &   &           &          \\
264.08 &  2.58$\pm$ 0.25 &  1.83$\pm$ 0.18 &  1.51$\pm$ 0.16 &  1.60$\pm$ 0.17 & ... & ... & ...\\
 ...   &  2.40 &  1.82 &  1.65 &  1.57 &  0 &  1.37e+00 &  7.12e$-$01\\
 ...   &  2.29 &  1.82 &  1.68 &  1.61 &  1 &  2.40e+00 &  4.95e$-$01\\
 ...   &  2.19 &  1.81 &  1.70 &  1.64 &  2 &  3.92e+00 &  2.71e$-$01\\
 ...   &  2.54 &  1.86 &  1.64 &  1.43 &  3 &  1.69e+00 &  6.39e$-$01\\
 ...   &  1.89 &  1.77 &  1.74 &  1.74 &  4 &  1.06e+01 &  1.43e$-$02\\
 ...   &  1.50 &  1.58 &  1.74 &  1.94 &  5 &  2.67e+01 &  6.83e$-$06\\

\enddata
\end{deluxetable}
\clearpage

\begin{deluxetable}{ccccccc}
\tablecolumns{7}
\tablewidth{0pt}
\tablecaption{Fits of predicted {\sl ULTRACAM} amplitudes to those observed for
  the pulsation modes of PG 0014+067 reported by J05}
\tablehead{
\colhead{Period} &
\colhead{$u'$} &
\colhead{$g'$} &
\colhead{$r'$} &
\colhead{$l$} &
\colhead{$\chi^2$} &
\colhead{$Q$}\\
\colhead{(s)} &
\colhead{(mmag)} &
\colhead{(mmag)} &
\colhead{(mmag)} &
\colhead{ } &
\colhead{ } &
\colhead{ }
}
 
\startdata

100.29 & 0.900$\pm$0.200 & 0.690$\pm$0.070 & 0.600$\pm$0.100 & ... & ... & ...\\
 ...   & 0.864 & 0.674 & 0.648 &  0 &  3.20e$-$01 &  8.52e$-$01\\
 ...   & 0.854 & 0.674 & 0.650 &  1 &  3.55e$-$01 &  8.37e$-$01\\
 ...   & 0.842 & 0.675 & 0.650 &  2 &  3.87e$-$01 &  8.24e$-$01\\
 ...   & 1.200 & 0.662 & 0.497 &  3 &  3.47e+00 &  1.77e$-$01\\
 ...   & 0.766 & 0.676 & 0.667 &  4 &  9.35e$-$01 &  6.27e$-$01\\
 ...   & 0.676 & 0.661 & 0.710 &  5 &  2.63e+00 &  2.68e$-$01\\
  &       &       &       &   &           &          \\
139.14 & 0.800$\pm$0.200 & 0.710$\pm$0.080 & 0.800$\pm$0.100 & ... & ... & ...\\
 ...   & 0.952 & 0.736 & 0.707 &  0 &  1.56e+00 &  4.59e$-$01\\
 ...   & 0.934 & 0.738 & 0.711 &  1 &  1.36e+00 &  5.05e$-$01\\
 ...   & 0.912 & 0.741 & 0.715 &  2 &  1.19e+00 &  5.51e$-$01\\
 ...   & 1.162 & 0.734 & 0.565 &  3 &  8.89e+00 &  1.18e$-$02\\
 ...   & 0.822 & 0.744 & 0.738 &  4 &  5.77e$-$01 &  7.49e$-$01\\
 ...   & 0.728 & 0.725 & 0.794 &  5 &  1.69e$-$01 &  9.19e$-$01\\
  &       &       &       &   &           &          \\
140.98 & 1.200$\pm$0.300 & 1.100$\pm$0.100 & 1.000$\pm$0.100 & ... & ... & ...\\
 ...   & 1.370 & 1.060 & 1.017 &  0 &  5.13e$-$01 &  7.74e$-$01\\
 ...   & 1.342 & 1.061 & 1.022 &  1 &  4.25e$-$01 &  8.08e$-$01\\
 ...   & 1.306 & 1.062 & 1.024 &  2 &  3.29e$-$01 &  8.48e$-$01\\
 ...   & 1.754 & 1.113 & 0.857 &  3 &  5.47e+00 &  6.50e$-$02\\
 ...   & 1.166 & 1.056 & 1.048 &  4 &  4.34e$-$01 &  8.05e$-$01\\
 ...   & 1.010 & 1.007 & 1.103 &  5 &  2.33e+00 &  3.12e$-$01\\
  &       &       &       &   &           &          \\
141.01 & 5.900$\pm$0.300 & 5.030$\pm$0.090 & 4.600$\pm$0.100 & ... & ... & ...\\
 ...   & 6.336 & 4.900 & 4.702 &  0 &  5.24e+00 &  7.28e$-$02\\
 ...   & 6.202 & 4.902 & 4.721 &  1 &  4.51e+00 &  1.05e$-$01\\
 ...   & 6.038 & 4.909 & 4.735 &  2 &  3.83e+00 &  1.47e$-$01\\
 ...   & 8.078 & 5.127 & 3.948 &  3 &  9.63e+01 &  1.20e$-$21\\
 ...   & 5.390 & 4.883 & 4.845 &  4 &  1.16e+01 &  3.10e$-$03\\
 ...   & 4.690 & 4.675 & 5.122 &  5 &  5.91e+01 &  1.50e$-$13\\
  &       &       &       &   &           &          \\
141.06 & 4.500$\pm$0.200 & 3.200$\pm$0.070 & 3.040$\pm$0.090 & ... & ... & ...\\
 ...   & 4.166 & 3.222 & 3.092 &  0 &  3.22e+00 &  2.00e$-$01\\
 ...   & 4.082 & 3.226 & 3.108 &  1 &  5.07e+00 &  7.91e$-$02\\
 ...   & 3.976 & 3.233 & 3.118 &  2 &  7.83e+00 &  2.00e$-$02\\
 ...   & 5.196 & 3.298 & 2.540 &  3 &  4.49e+01 &  1.74e$-$10\\
 ...   & 3.564 & 3.229 & 3.204 &  4 &  2.54e+01 &  3.07e$-$06\\
 ...   & 3.128 & 3.118 & 3.416 &  5 &  6.59e+01 &  4.90e$-$15\\
  &       &       &       &   &           &          \\
146.50 & 4.000$\pm$0.200 & 3.030$\pm$0.060 & 2.680$\pm$0.080 & ... & ... & ...\\
 ...   & 3.834 & 2.963 & 2.842 &  0 &  6.05e+00 &  4.85e$-$02\\
 ...   & 3.750 & 2.964 & 2.855 &  1 &  7.53e+00 &  2.32e$-$02\\
 ...   & 3.646 & 2.969 & 2.863 &  2 &  9.43e+00 &  8.97e$-$03\\
 ...   & 4.766 & 3.060 & 2.361 &  3 &  3.08e+01 &  2.04e$-$07\\
 ...   & 3.258 & 2.959 & 2.938 &  4 &  2.55e+01 &  2.86e$-$06\\
 ...   & 2.858 & 2.854 & 3.133 &  5 &  7.32e+01 &  1.25e$-$16\\
  &       &       &       &   &           &          \\
150.47 & 0.900$\pm$0.200 & 0.690$\pm$0.070 & 0.620$\pm$0.090 & ... & ... & ...\\
 ...   & 0.876 & 0.677 & 0.649 &  0 &  1.54e$-$01 &  9.26e$-$01\\
 ...   & 0.858 & 0.678 & 0.653 &  1 &  2.08e$-$01 &  9.01e$-$01\\
 ...   & 0.834 & 0.680 & 0.656 &  2 &  2.87e$-$01 &  8.66e$-$01\\
 ...   & 1.076 & 0.696 & 0.538 &  3 &  1.62e+00 &  4.45e$-$01\\
 ...   & 0.746 & 0.679 & 0.674 &  4 &  9.79e$-$01 &  6.13e$-$01\\
 ...   & 0.654 & 0.654 & 0.719 &  5 &  2.98e+00 &  2.25e$-$01\\
  &       &       &       &   &           &          \\
150.78 & 0.800$\pm$0.100 & 0.570$\pm$0.070 & 0.600$\pm$0.100 & ... & ... & ...\\
 ...   & 0.774 & 0.598 & 0.573 &  0 &  2.97e$-$01 &  8.62e$-$01\\
 ...   & 0.762 & 0.602 & 0.580 &  1 &  3.99e$-$01 &  8.19e$-$01\\
 ...   & 0.746 & 0.608 & 0.587 &  2 &  6.06e$-$01 &  7.39e$-$01\\
 ...   & 0.880 & 0.570 & 0.440 &  3 &  3.20e+00 &  2.02e$-$01\\
 ...   & 0.684 & 0.622 & 0.618 &  4 &  1.94e+00 &  3.80e$-$01\\
 ...   & 0.618 & 0.618 & 0.679 &  5 &  4.41e+00 &  1.10e$-$01\\
  &       &       &       &   &           &          \\
154.94 & 0.700$\pm$0.200 & 0.670$\pm$0.060 & 0.780$\pm$0.080 & ... & ... & ...\\
 ...   & 0.910 & 0.703 & 0.674 &  0 &  3.16e+00 &  2.05e$-$01\\
 ...   & 0.890 & 0.704 & 0.678 &  1 &  2.86e+00 &  2.39e$-$01\\
 ...   & 0.864 & 0.705 & 0.680 &  2 &  2.58e+00 &  2.76e$-$01\\
 ...   & 1.096 & 0.714 & 0.552 &  3 &  1.26e+01 &  1.87e$-$03\\
 ...   & 0.776 & 0.707 & 0.702 &  4 &  1.47e+00 &  4.81e$-$01\\
 ...   & 0.686 & 0.687 & 0.756 &  5 &  1.75e$-$01 &  9.16e$-$01\\
  &       &       &       &   &           &          \\
168.80 & 0.600$\pm$0.200 & 0.410$\pm$0.070 & 0.400$\pm$0.100 & ... & ... & ...\\
 ...   & 0.542 & 0.418 & 0.401 &  0 &  9.74e$-$02 &  9.52e$-$01\\
 ...   & 0.530 & 0.419 & 0.403 &  1 &  1.41e$-$01 &  9.32e$-$01\\
 ...   & 0.514 & 0.420 & 0.405 &  2 &  2.10e$-$01 &  9.00e$-$01\\
 ...   & 0.644 & 0.427 & 0.331 &  3 &  5.85e$-$01 &  7.46e$-$01\\
 ...   & 0.460 & 0.421 & 0.418 &  4 &  5.47e$-$01 &  7.61e$-$01\\
 ...   & 0.406 & 0.408 & 0.450 &  5 &  1.19e+00 &  5.52e$-$01\\

\enddata
\end{deluxetable}
\clearpage

\begin{deluxetable}{lccc}
\tablecolumns{4}
\tablewidth{0pt}
\tablecaption{Summary of partial mode identification in pulsating sdB stars}
\tablehead{
\colhead{Object} &
\colhead{Period} &
\colhead{Possible $l$ identification} &
\colhead{Mode discrimination}\\
\colhead{ } &
\colhead{(s)} &
\colhead{(0$-$5)} &
\colhead{ } 
}
 
\startdata

KPD 2109+4401 (K98) & 182.42 & 0,1,2 & partial\\
... & 184.72 & 0,1,2,4 & partial\\
... & 184.75 & 2,4 & partial\\
... & 191.85 & 1,2,4 & partial\\
... & 196.31 & 0,1,2 & partial\\
... & 196.69 & 0,1,2,3,4,5 & no\\
... & 198.19 & 2,4 & partial\\
 & & & \\
PG 1618+563B (S00) & 139.30 & 0,1,2,3,4,5 & no\\
... & 143.90 & 0,1,2,3,4,5 & no\\
 & & & \\
PG 1605+072 (F03) & 440.51 & 0,1,2,3,4,5 & no\\
... & 475.82 & 0,1,2,4 & partial\\
... & 481.75 & 0,1,2 & partial\\
... & 503.70 & 0,1,2,4 & partial\\
... & 528.70 & 0,1,2,3,4,5 & no\\
 & & & \\
HS 0039+4302 (J04) & 134.44 & 0,1,2,3,4,5 & no\\
... & 180.13 & 0,1,2,4,5 & partial\\
... & 181.89 & 0,1,2 & partial\\
... & 182.79 & 0,1,2,4 & partial\\
... & 192.58 & 0,1,2 & partial\\
... & 234.11 & 0,1,2 & partial\\
 & & & \\
BAL 090100001 (O05) & 351.70 & no acceptable fit & no\\
... & 356.30 & no acceptable fit & no\\
 & & & \\
BAL 090100001 (B05) & 264.08 & 0,1,2,3,4 & partial\\
... & 264.82 & 0,1,2,3 & partial\\
... & 349.83 & 0,1,2,3,4,5 & no\\
... & 350.18 & 0,1,2,3,4 & partial\\
... & 350.39 & 0,1,2,3,4 & partial\\
... & 353.81 & 1,2,4 & partial\\
... & 354.01 & 1,2 & partial\\
... & 354.20 & 1,2 & partial\\
... & 356.19 & 0,1 & partial\\
 & & & \\
PG 0014+067 (J05) & 100.29 & 0,1,2,3,4,5 & no\\
... & 139.14 & 0,1,2,3,4,5 & no\\
... & 140.98 & 0,1,2,3,4,5 & no\\
... & 141.01 & 0,1,2,4 & partial\\
... & 141.06 & 0,1,2 & partial\\
... & 146.50 & 0,1,2 & partial\\
... & 150.47 & 0,1,2,3,4,5 & no\\
... & 150.78 & 0,1,2,3,4,5 & no\\
... & 154.94 & 0,1,2,3,4,5 & no\\
... & 168.80 & 0,1,2,3,4,5 & no\\
\enddata
\end{deluxetable}
\clearpage


\clearpage
\centerline{\bf{FIGURE CAPTIONS}}

\noindent Fig. 1 ---  Fits to the $U$, $B$, $V$, and $R$ pulsational
amplitudes observed for the 182.42 s mode of KPD 2109+4401 by K98. The
predicted amplitude-wavelength behaviors of modes with $l = 0$ to $l =
5$ have been fit to the observed values using a least-squares
procedure. The curves with $l = 0$, $l = 1$, and $l = 2$ provide
acceptable fits according to the quality-of-fit $Q$ quantity (see
text). In contrast, the curves with $l = 3$, $l = 4$, and $l = 5$ do not
provide viable fits and, consequently, these possible $l$
identifications must be rejected for that mode.

\noindent Fig. 2 ---  Similar to Fig. 1, but for the mode with a
period of 196.31 s. Here the acceptable fits are again for values of the
degree index  $l = 0$, $l = 1$, and $l = 2$. 

\noindent Fig. 3 ---  Similar to Fig. 1, but for the mode with a
period of 198.19 s. Here the only acceptable fit from a formal point of
view is the one corresponding to $l = 4$. 

\noindent Fig. 4 ---  Fits to the $U$, $B$, and $V$ pulsational
amplitudes observed for the 139.3 s mode of PG 1618+563B by S00. No mode
discrimination is possible here.

\noindent Fig. 5 ---  Fits to the {\sl BUSCA} ($UV$, $B$, $R$, and
$NIR$) pulsational amplitudes observed for the 475.82 s mode of PG
1605+072 by F03. Partial mode discrimination is possible here, with the
values $l = 3$ and $l = 5$ excluded.

\noindent Fig. 6 ---  Similar to Fig. 5, but for the mode with a
period of 481.75 s. Here the acceptable fits are for values of the
degree index  $l = 0$, $l = 1$, and $l = 2$. 

\noindent Fig. 7 ---  Similar to Fig. 5, but for the mode with a
period of 503.70 s. Partial mode discrimination is possible, with 
the values $l = 3$ and $l = 5$ excluded.

\noindent Fig. 8 ---  Fits to the {\sl ULTRACAM} ($u'$, $g'$, and 
$r'$) pulsational amplitudes observed for the 181.89 s mode of HS
0039+4302 by J04. The acceptable values of $l$ for that mode are $l =
0$, $l = 1$, and $l =2$.

\noindent Fig. 9 ---  Similar to Fig. 8, but for the mode with a
period of 192.58 s. Here the acceptable fits are for values of the
degree index  $l = 0$, $l = 1$, and $l = 2$. 

\noindent Fig. 10 ---  Similar to Fig. 8, but for the mode with a
period of 243.11 s. The acceptable fits are for values of the
degree index  $l = 0$, $l = 1$, and $l = 2$.

\noindent Fig. 11 ---  Fits to the $u$, $v$, $b$ and $y$ pulsational
amplitudes observed for the 356.3 s mode of BAL 090100001 by O05. No mode
discrimination is possible here.

\noindent Fig. 12 ---  Fits to the $U$, $B$, $V$, and $R$ pulsational
amplitudes observed for the 356.19 s mode of BAL 090100001 by B05.
There is no acceptable formal fit, but it is clear that the mode must
have a degree index $l = 0$ or $l = 1$.

\noindent Fig. 13 ---  Similar to Fig. 12, but for the mode with a
period of 354.20 s. Formally, the only acceptable ($Q > 0.001$) fit is
that with $l =2$, although the solution with $l = 1$ should not be
discarded since the data points fall in between those two model curves. 

\noindent Fig. 14 ---  Similar to Fig. 12, but for the mode with a
period of 354.01 s. The acceptable fits are for values of the
degree index  $l = 1$, and $l = 2$.

\noindent Fig. 15 ---  Similar to Fig. 12, but for the mode with a
period of 353.81 s. The acceptable fits are for values of the
degree index  $l = 1$, $l = 2$, and $l = 4$.

\noindent Fig. 16 ---  Fits to the {\sl ULTRACAM} ($u'$, $g'$, and 
$r'$) pulsational amplitudes observed for the 141.01 s mode of PG 
0014+067 by J05. The acceptable values of $l$ for that mode are $l =
0$, $l = 1$, $l = 2$, and $l =4$.

\noindent Fig. 17 ---  Similar to Fig. 16, but for the mode with a
period of 141.06 s. Here the acceptable fits are for values of the
degree index  $l = 0$, $l = 1$, and $l = 2$. 

\noindent Fig. 18 ---  Similar to Fig. 16, but for the mode with a
period of 146.50 s. The acceptable fits are for values of the
degree index  $l = 0$, $l = 1$, and $l = 2$.

\clearpage
\begin{figure}[p]
\plotone{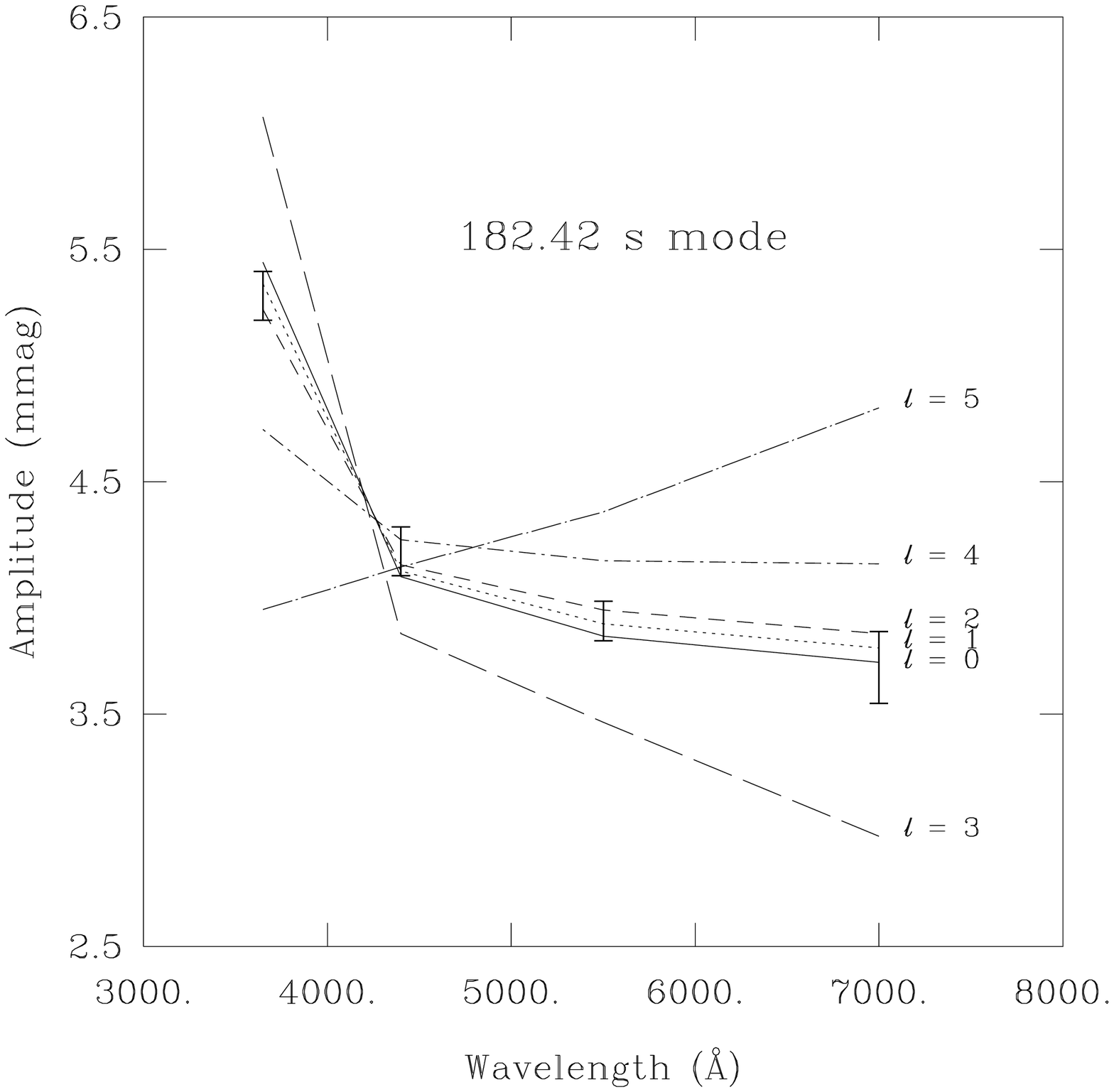}
\begin{flushright}
Figure 1
\end{flushright}
\end{figure}

\clearpage
\begin{figure}[p]
\plotone{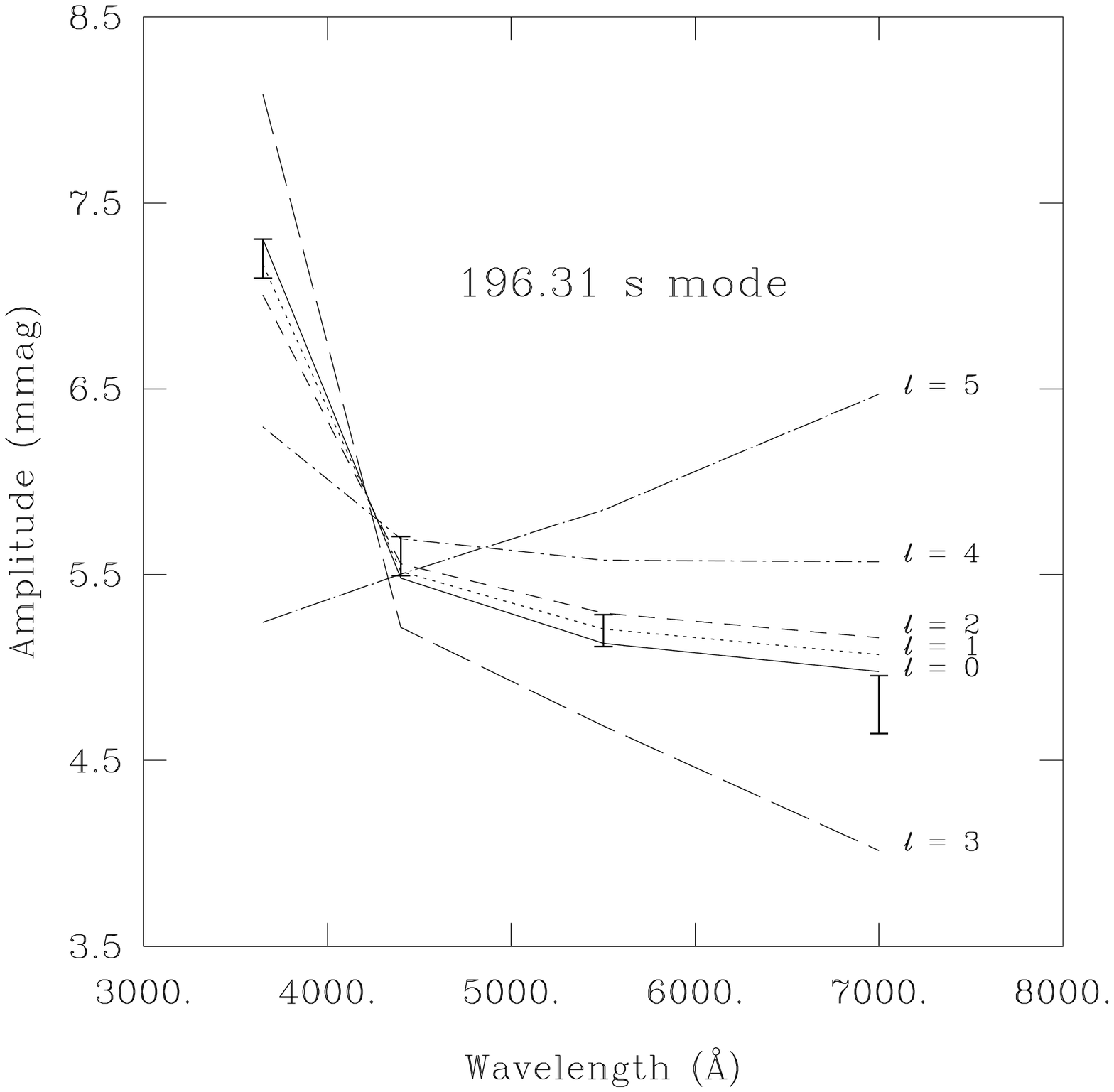}
\begin{flushright}
Figure 2
\end{flushright}
\end{figure}

\clearpage
\begin{figure}[p]
\plotone{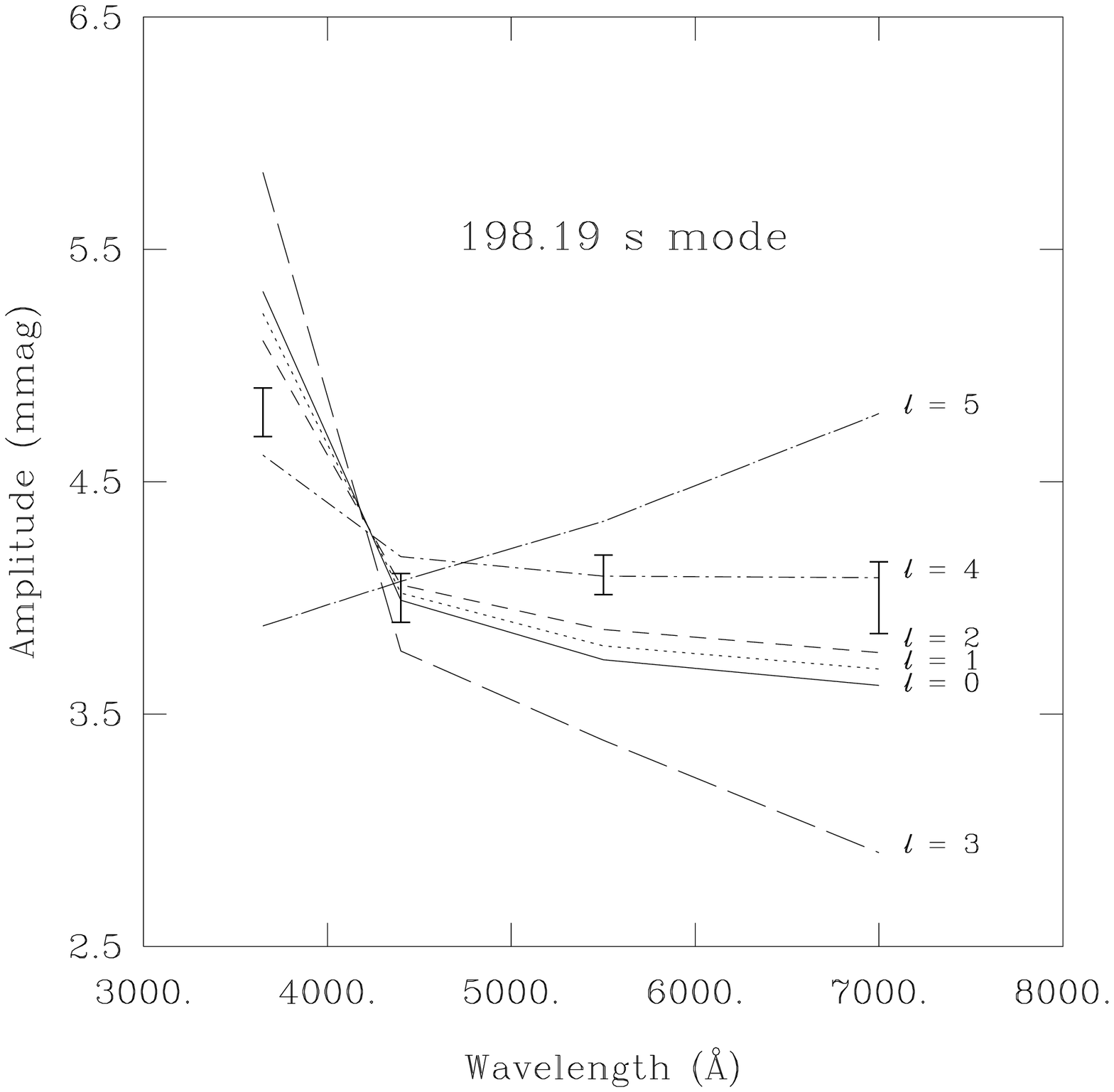}
\begin{flushright}
Figure 3
\end{flushright}
\end{figure}

\clearpage
\begin{figure}[p]
\plotone{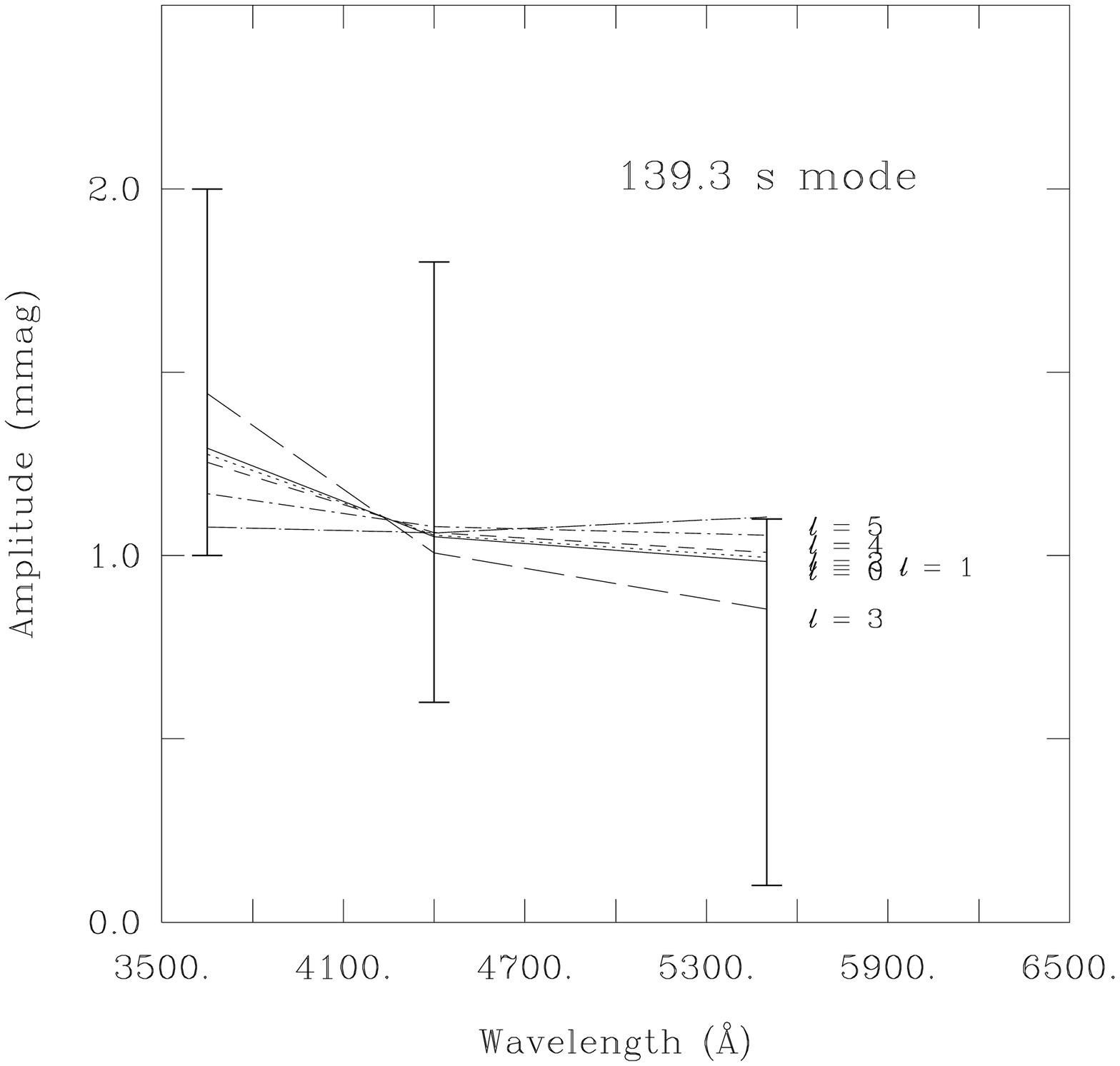}
\begin{flushright}
Figure 4
\end{flushright}
\end{figure}

\clearpage
\begin{figure}[p]
\plotone{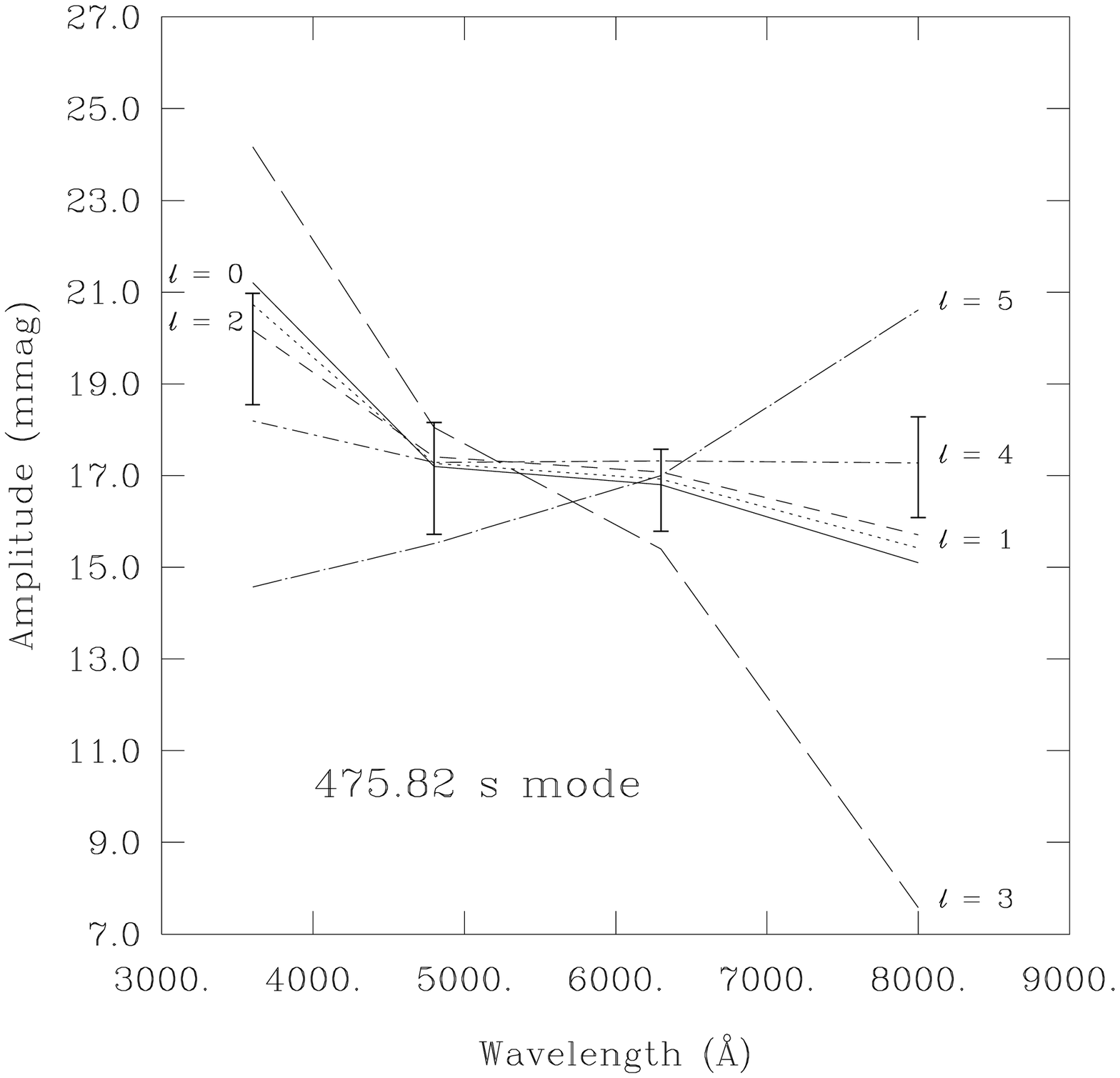}
\begin{flushright}
Figure 5
\end{flushright}
\end{figure}

\clearpage
\begin{figure}[p]
\plotone{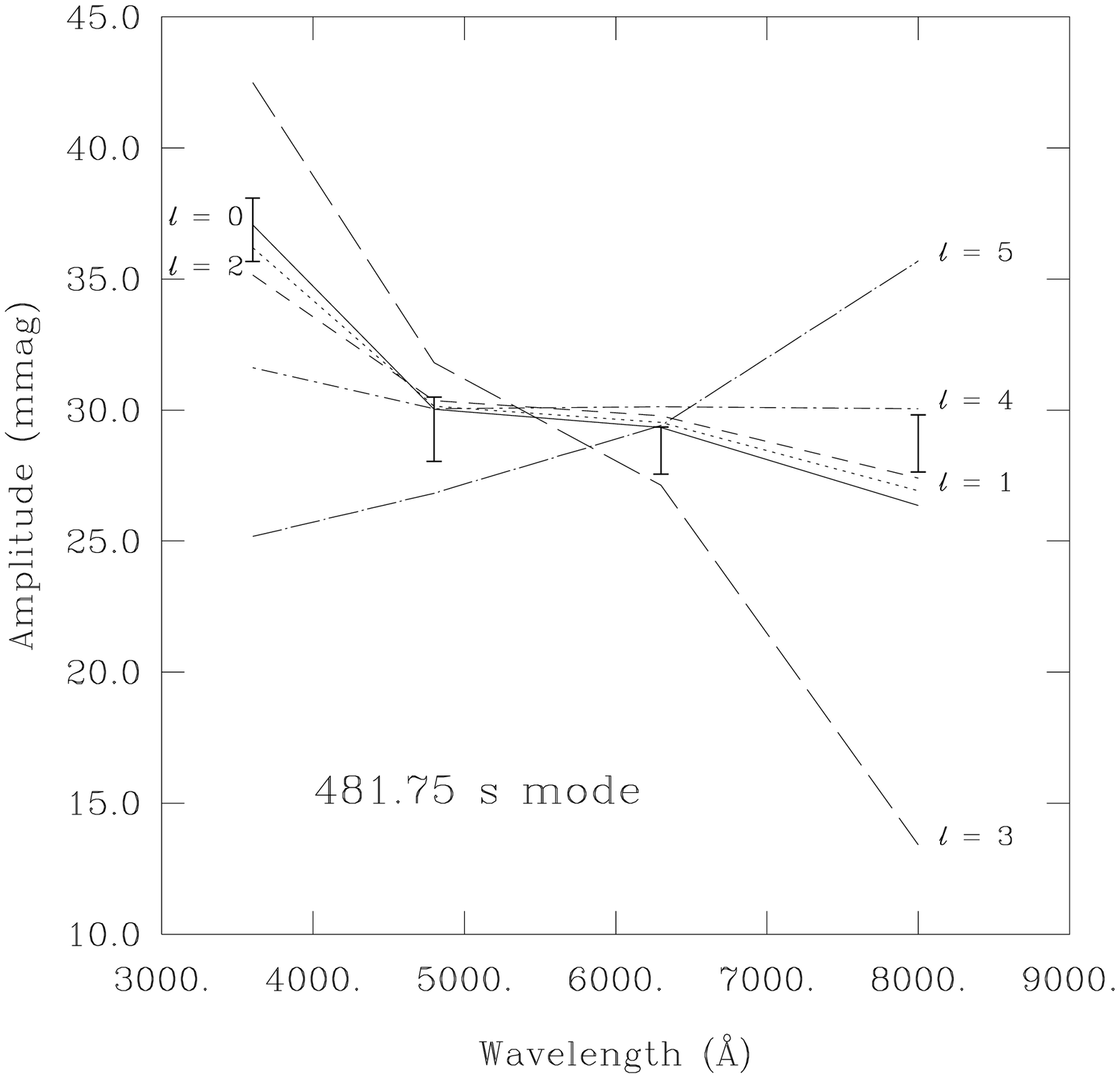}
\begin{flushright}
Figure 6
\end{flushright}
\end{figure}

\clearpage
\begin{figure}[p]
\plotone{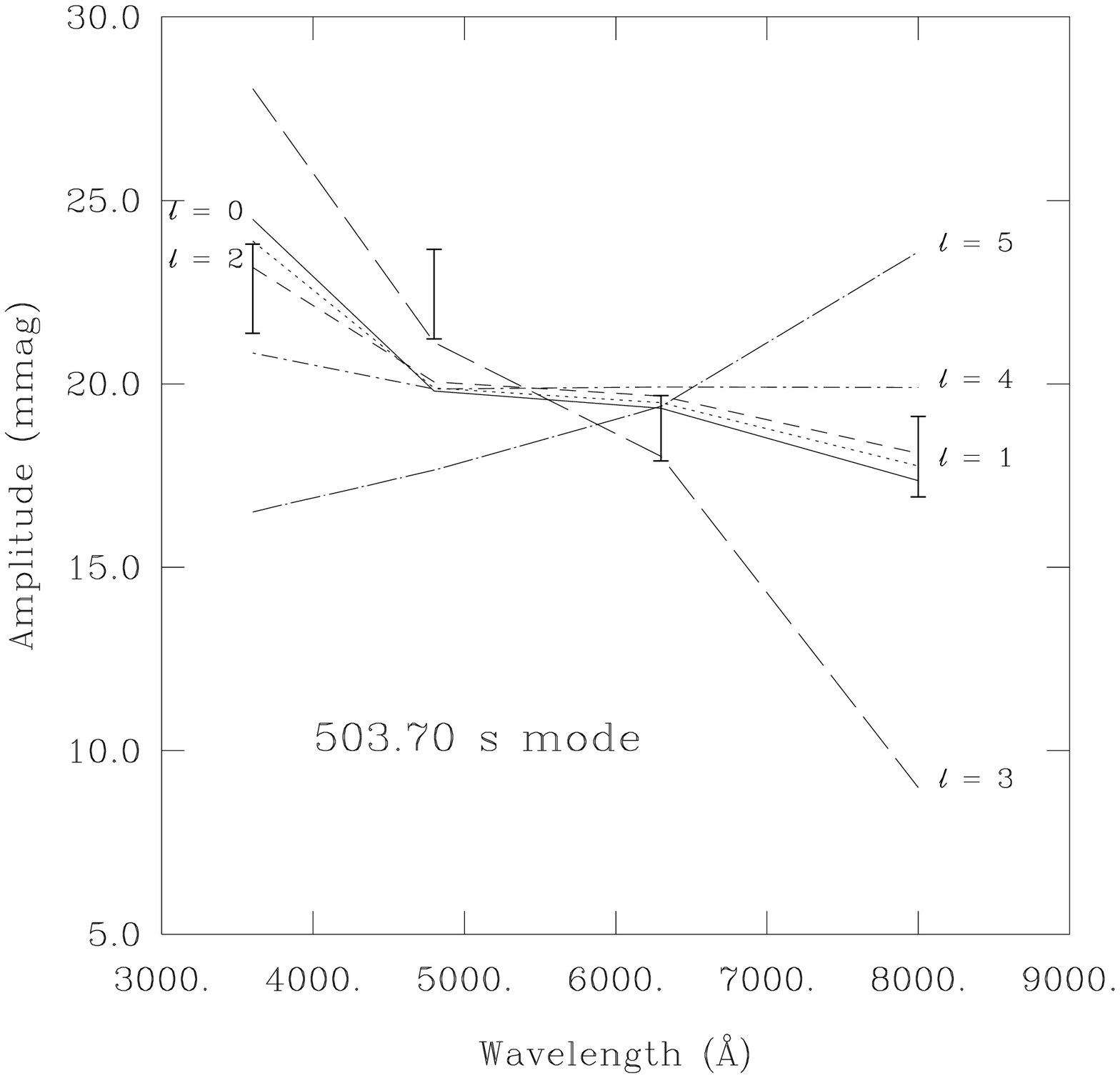}
\begin{flushright}
Figure 7
\end{flushright}
\end{figure}

\clearpage
\begin{figure}[p]
\plotone{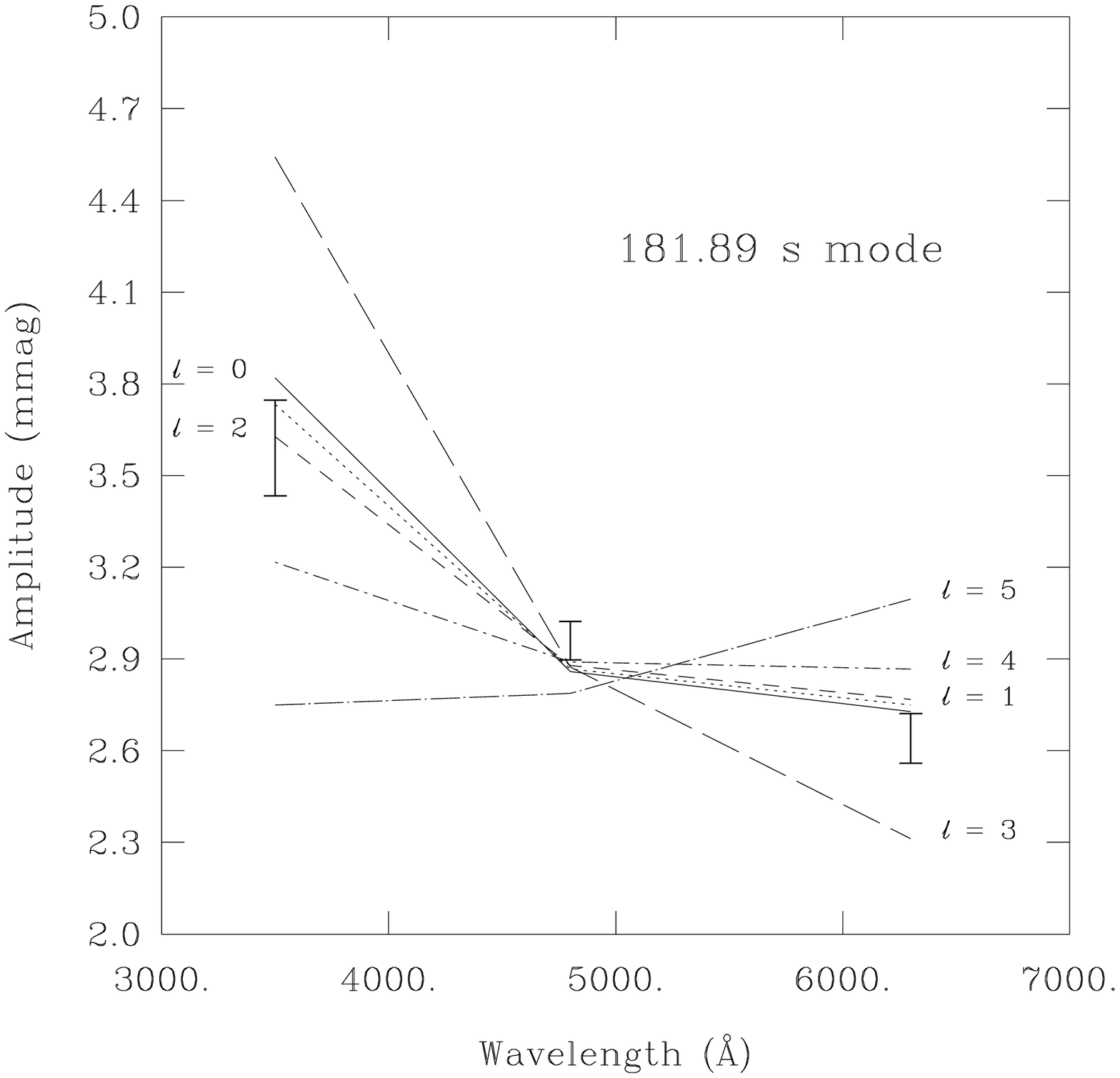}
\begin{flushright}
Figure 8
\end{flushright}
\end{figure}

\clearpage
\begin{figure}[p]
\plotone{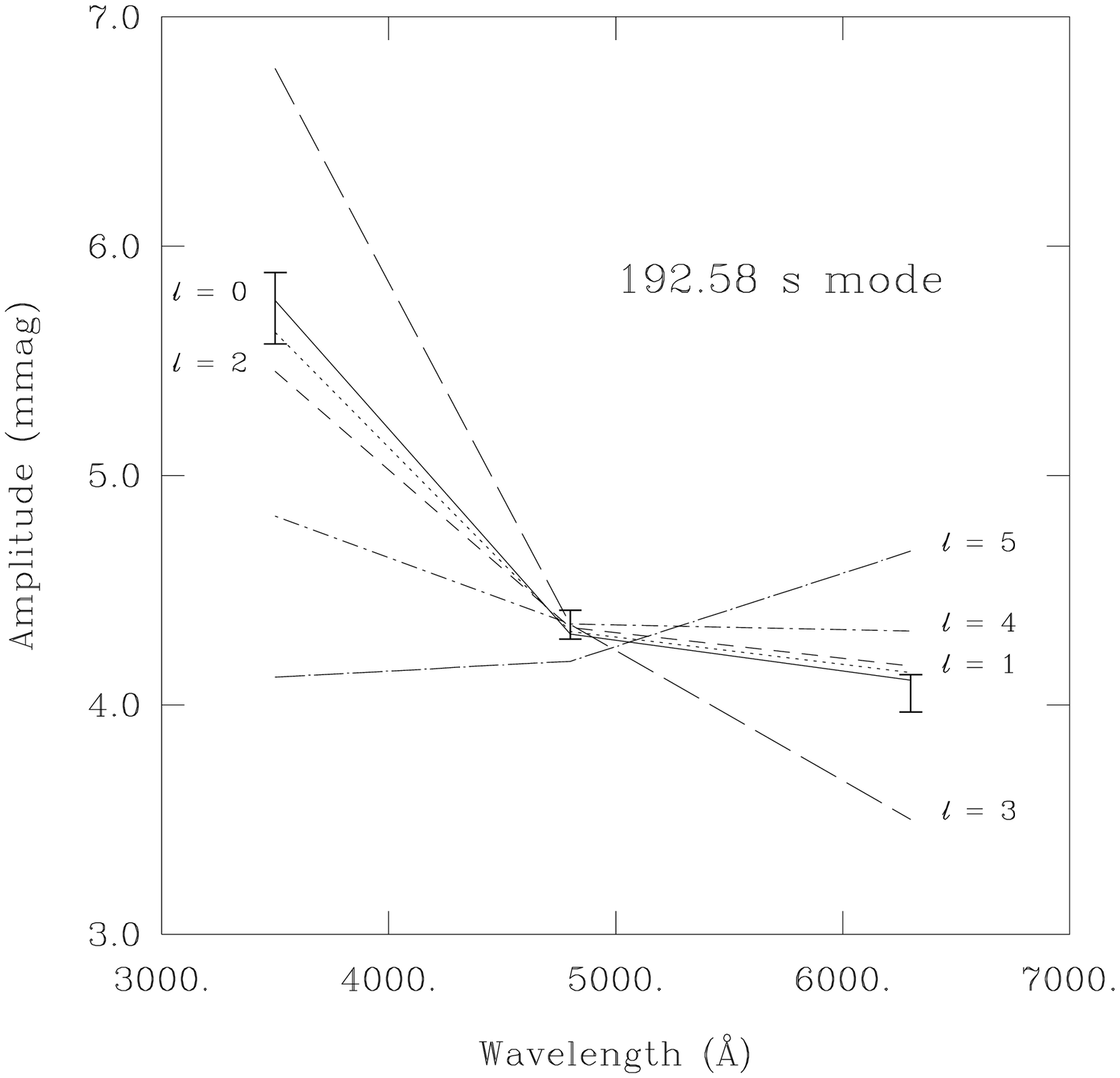}
\begin{flushright}
Figure 9
\end{flushright}
\end{figure}

\clearpage
\begin{figure}[p]
\plotone{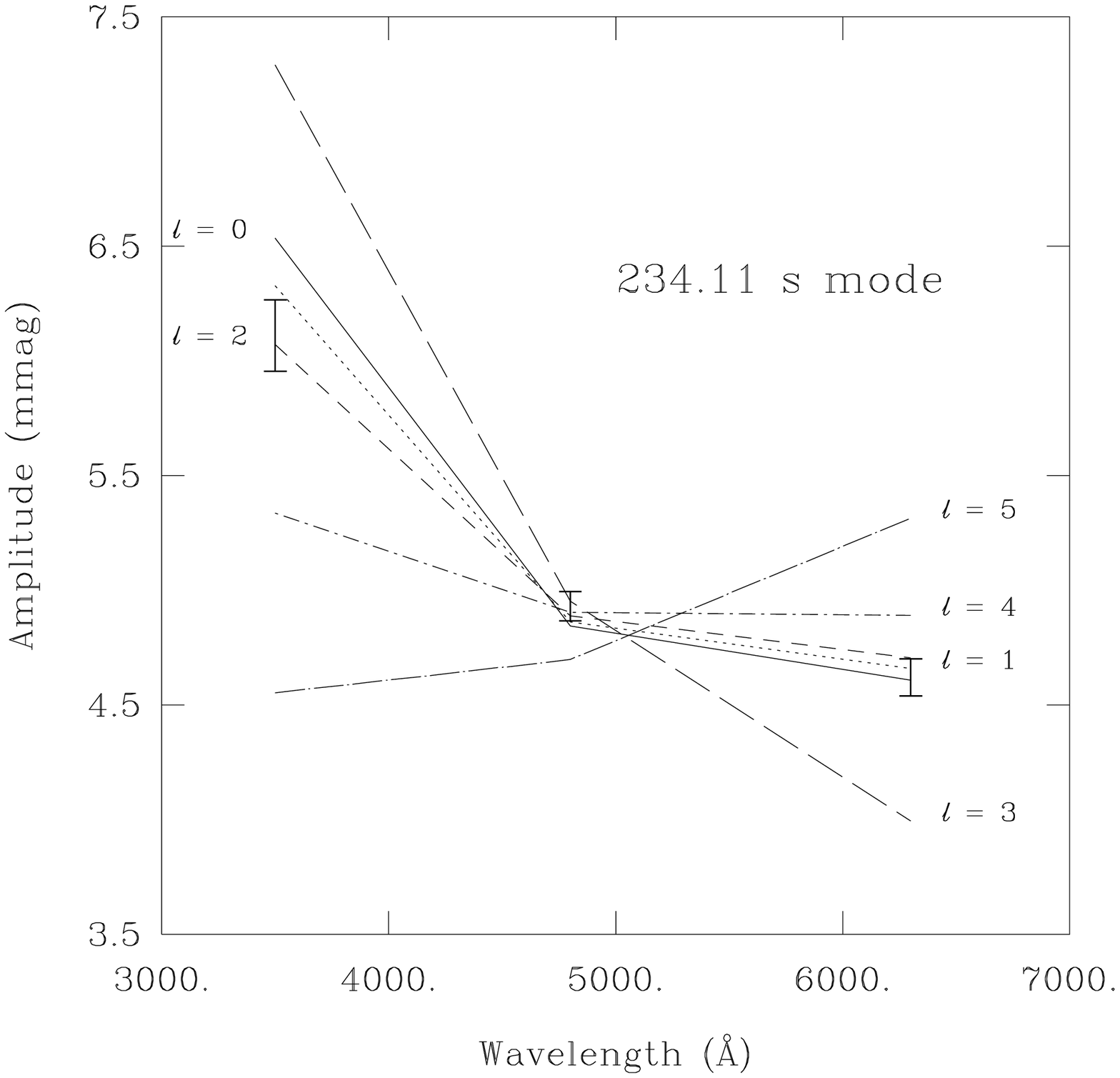}
\begin{flushright}
Figure 10
\end{flushright}
\end{figure}

\clearpage
\begin{figure}[p]
\plotone{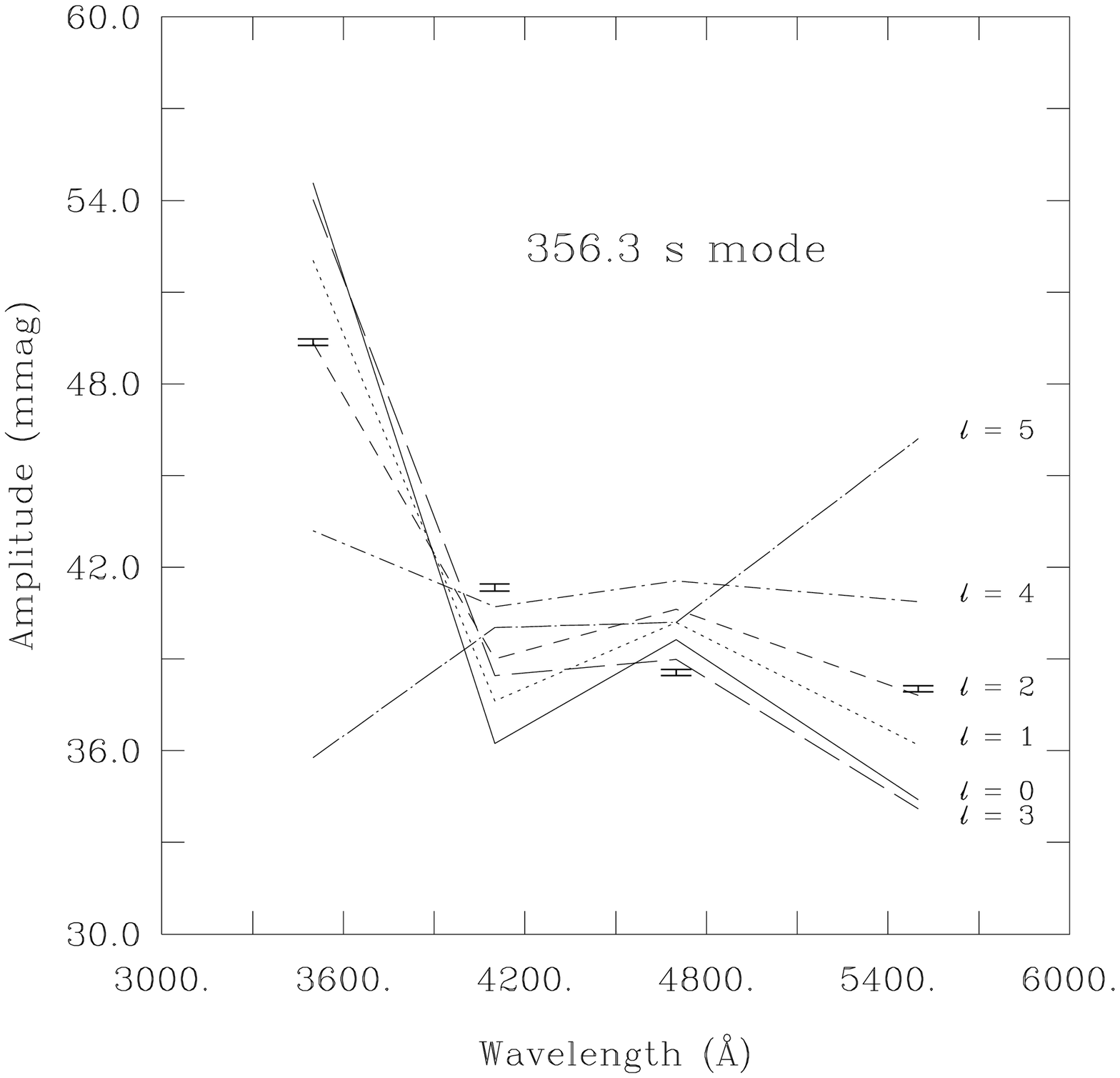}
\begin{flushright}
Figure 11
\end{flushright}
\end{figure}

\clearpage
\begin{figure}[p]
\plotone{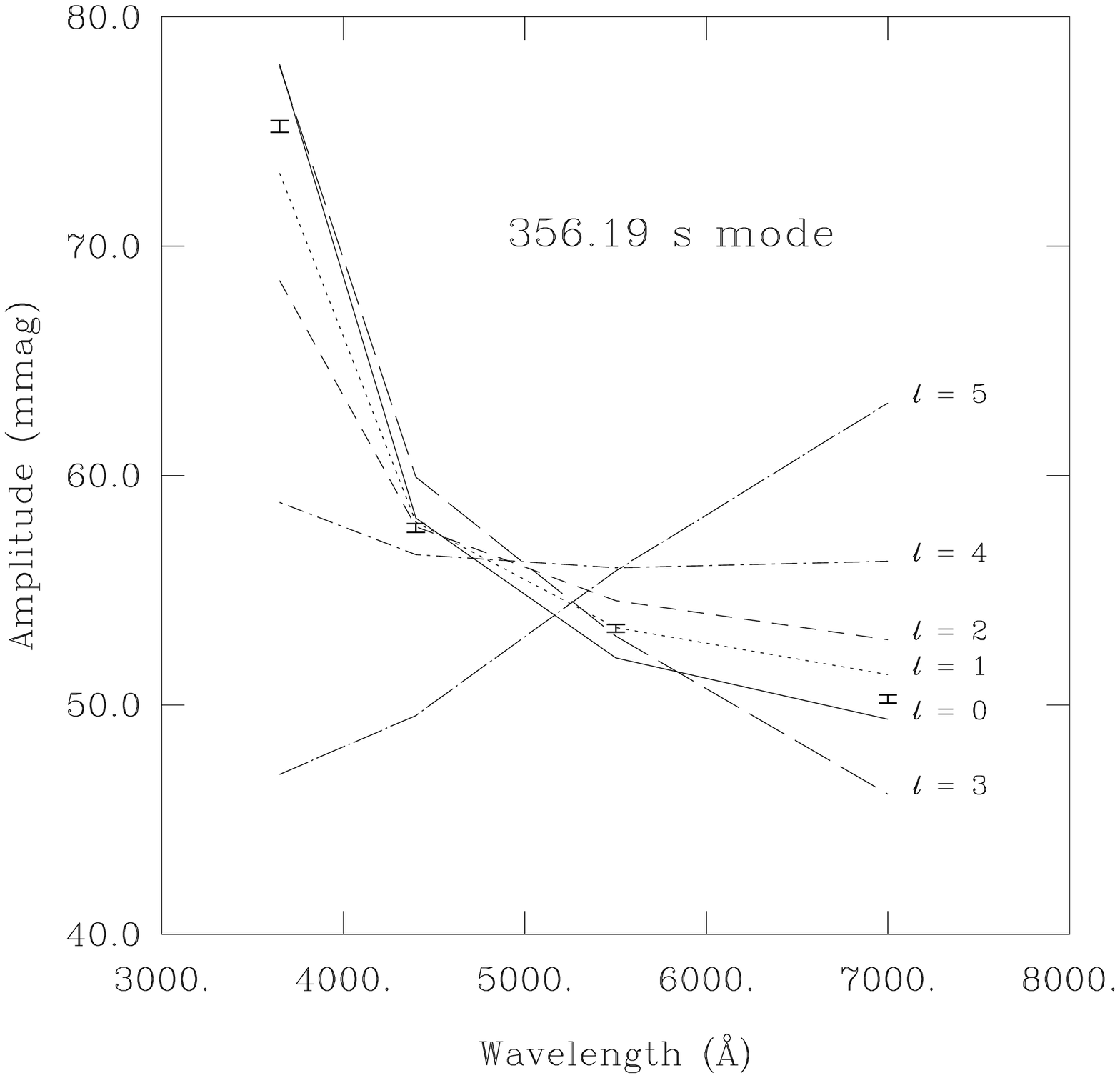}
\begin{flushright}
Figure 12
\end{flushright}
\end{figure}

\clearpage
\begin{figure}[p]
\plotone{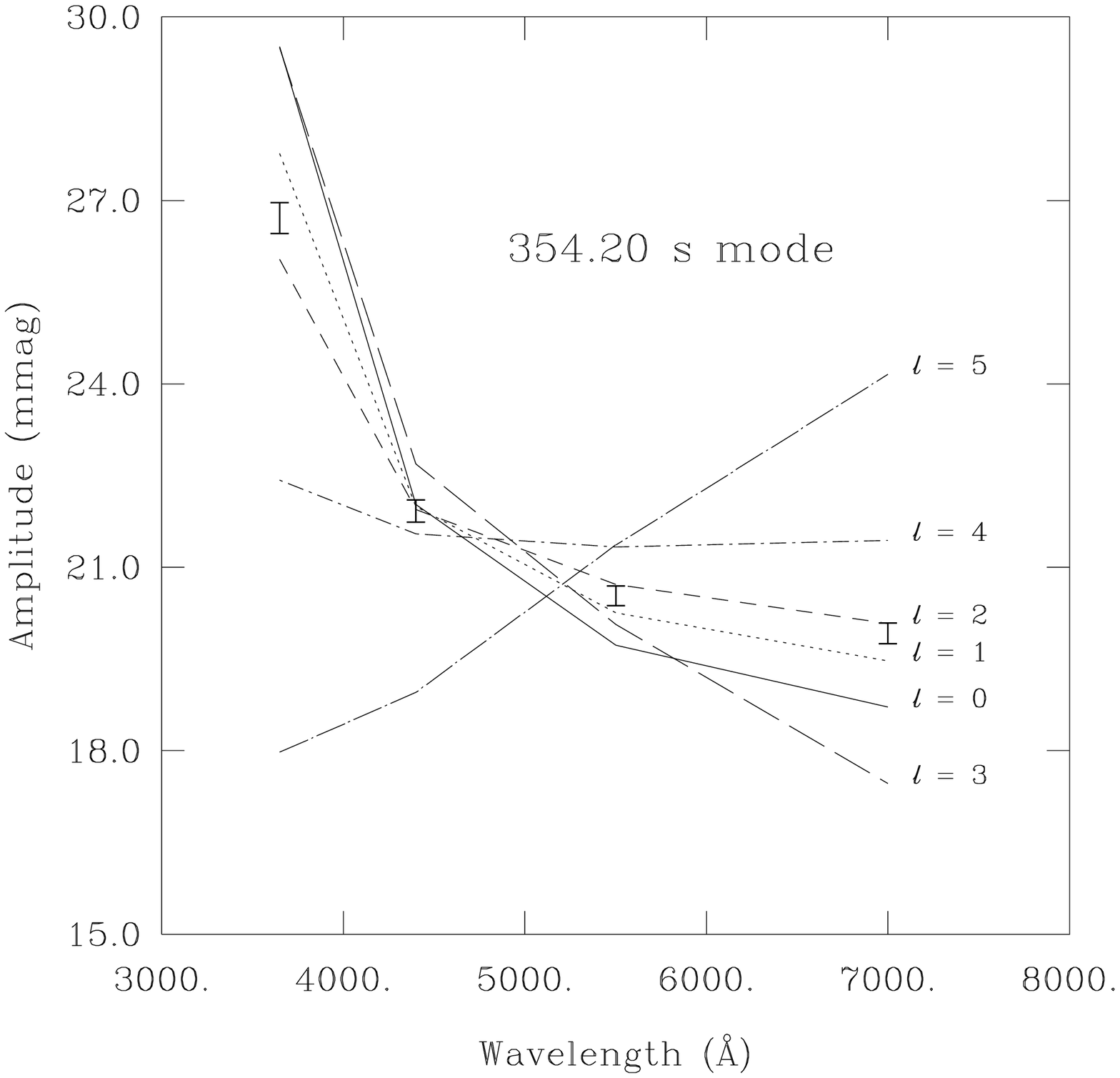}
\begin{flushright}
Figure 13
\end{flushright}
\end{figure}

\clearpage
\begin{figure}[p]
\plotone{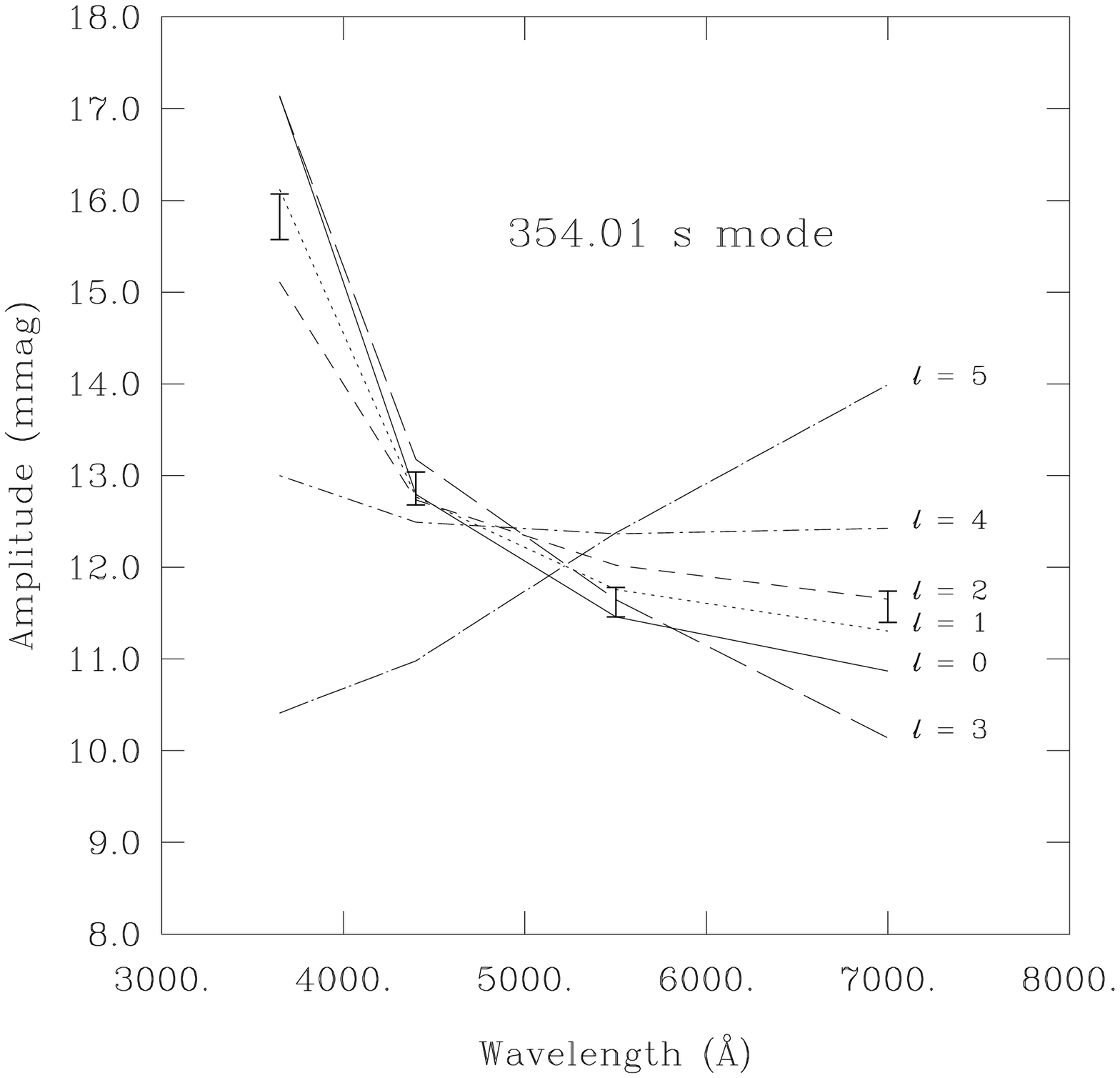}
\begin{flushright}
Figure 14
\end{flushright}
\end{figure}

\clearpage
\begin{figure}[p]
\plotone{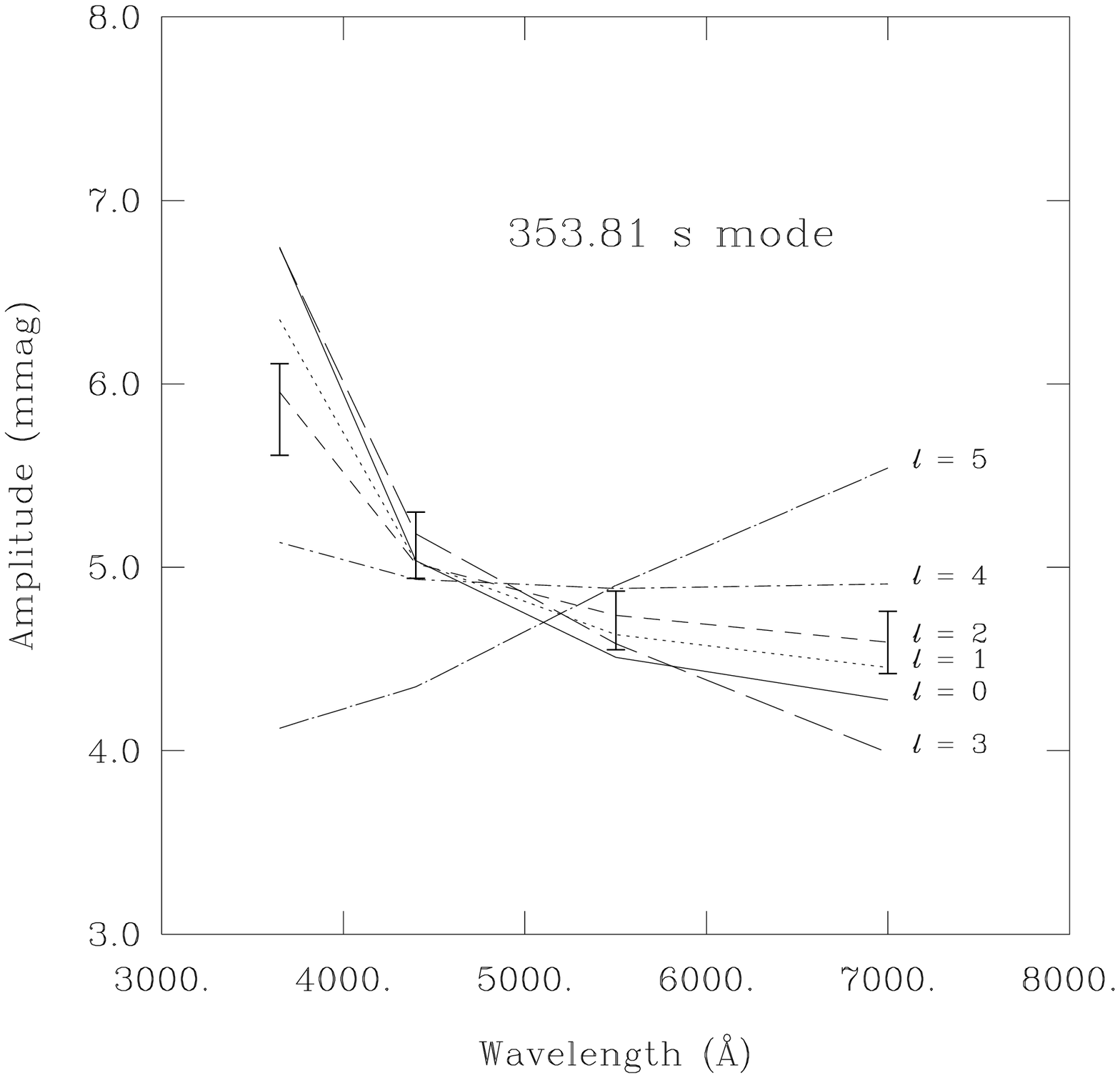}
\begin{flushright}
Figure 15
\end{flushright}
\end{figure}

\clearpage
\begin{figure}[p]
\plotone{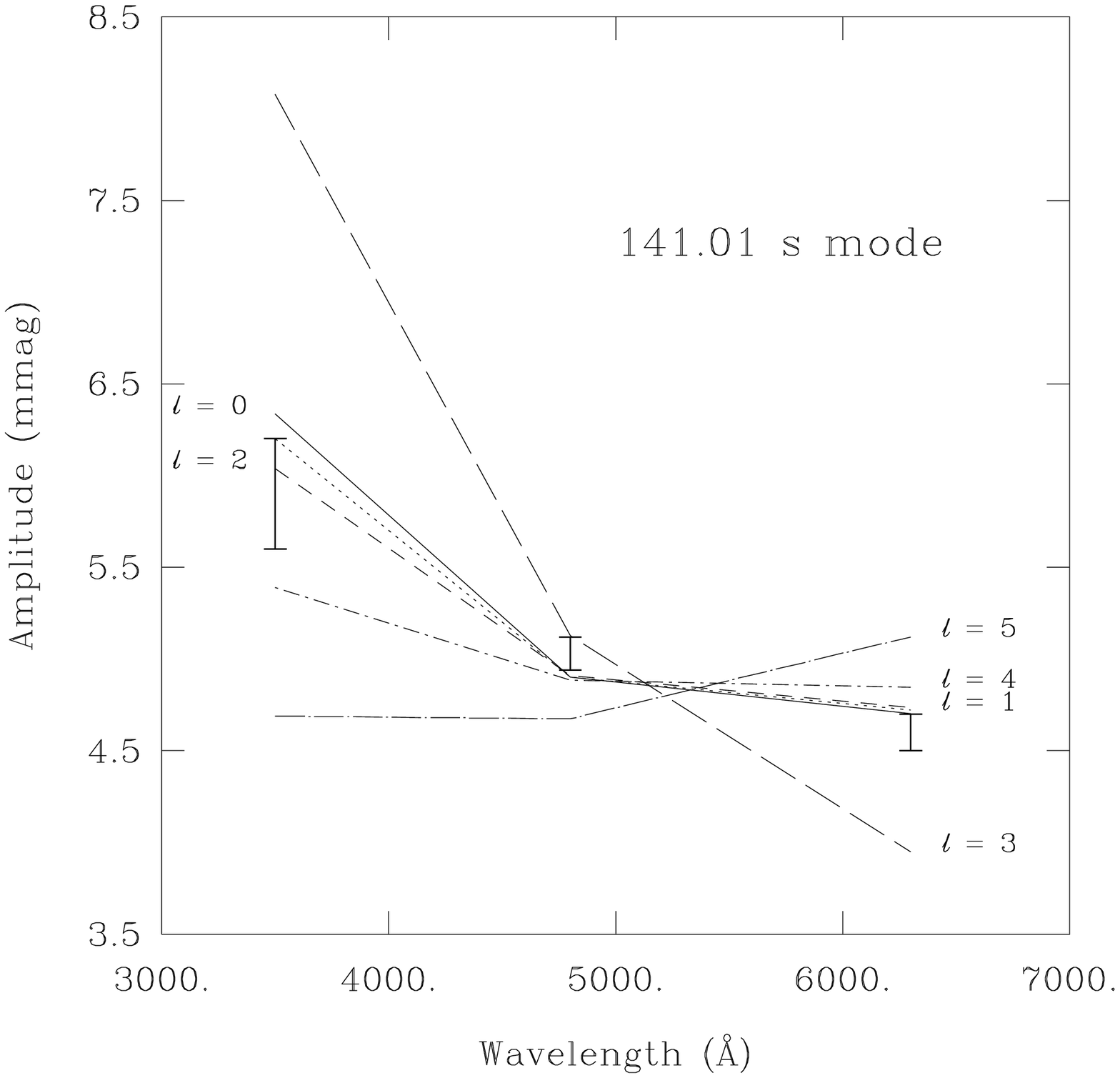}
\begin{flushright}
Figure 16
\end{flushright}
\end{figure}

\clearpage
\begin{figure}[p]
\plotone{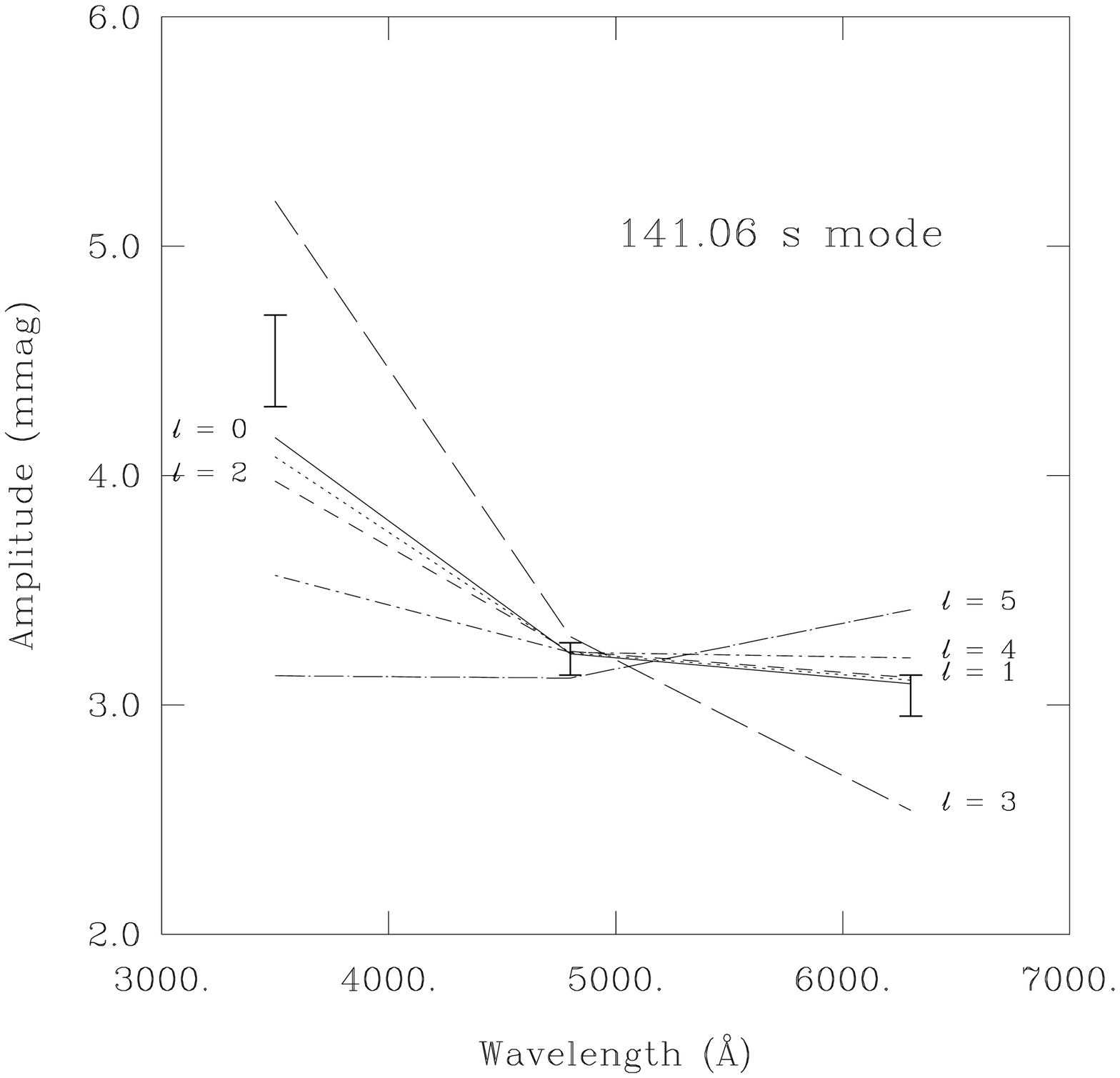}
\begin{flushright}
Figure 17
\end{flushright}
\end{figure}

\clearpage
\begin{figure}[p]
\plotone{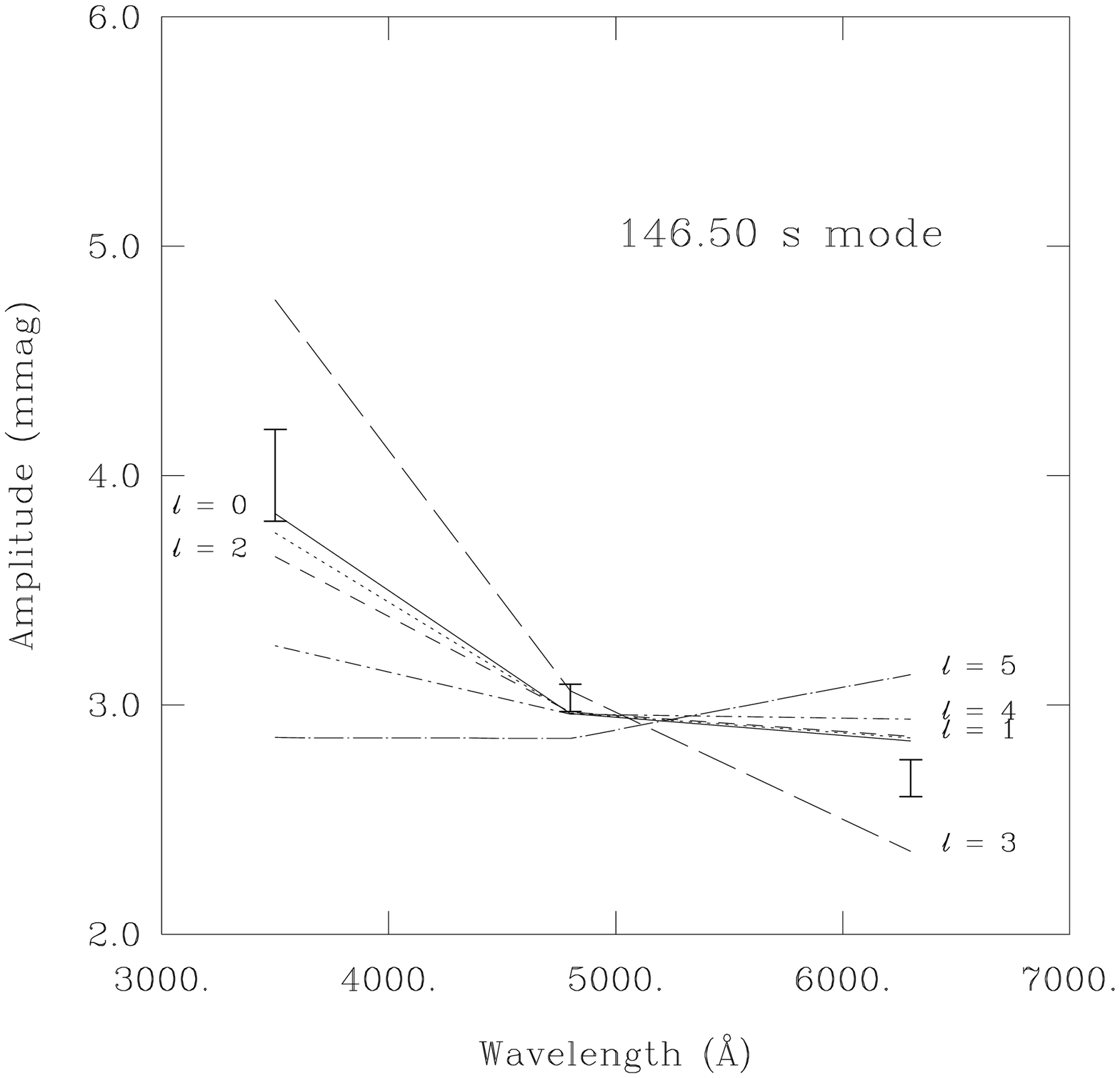}
\begin{flushright}
Figure 18
\end{flushright}
\end{figure}

\end{document}